Original Research Article

# Metabolic heat and moisture reuse with climate-adaptive control for energy-efficient black soldier fly bioconversion in modular container farms

Heat and moisture reuse in BSF farm

S. Saravanan[1,a], P. Sauerteig[1,b], M. Gold[2,3,c], S. Streif[1,4,d,*]

[1] Professorship Automatic Control and System Dynamics, Technische Universität Chemnitz, 09126 Chemnitz, Germany (e-mail: sriram.saravanan@etit.tu-chemnitz.de; philipp.sauerteig@etit.tu-chemnitz.de; stefan.streif@etit.tu-chemnitz.de).
[2] ETH Zurich, Sustainable Food Processing, Zurich, Switzerland
[3] Reploid Group AG, Wels, Austria (e-mail: gold@reploid.eu)
[4] Fraunhofer Institute for Molecular Biology and Applied Ecology IME, Department of Bioresources, Giessen, Germany.

*Corresponding author: stefan.streif@etit.tu-chemnitz.de.

## Abstract

Black soldier fly larvae (BSFL, *Hermetia illucens*) convert organic residues into feed ingredients and fertiliser. Climate control is a major contributor to the energy demand of container-based bioconversion. We developed a control-oriented dynamic model linking larval growth, substrate dynamics and container climate to heating, ventilation and air conditioning (HVAC). A two-level Model Predictive Control (MPC) strategy planned climate trajectories and tracked them through HVAC actuation. Four simulation studies examined controller selection, HVAC configuration, larval age pairing and inter-container coupling, using subsets of eight climate scenarios. Across the four scenarios tested for controller selection, dynamic MPC reduced specific energy consumption (SEC) by 31–61% and increased predicted dry biomass at harvest by up to 27.5% relative to single fixed setpoints. Under the tested controller settings, active cooling was required in the two hot scenarios, whereas dehumidification provided no additional operational benefit. Pairing larval ages four days apart within one container reduced SEC by up to 33% in the tested non-hot scenarios. In the $\mathrm{Cool}_{\mathrm{dry}}$ scenario, two age-separated containers connected by an actuated air duct produced 9% more biomass and achieved 3.9% lower SEC than two independent containers with mixed ages, despite using 4.6% more total HVAC energy. The simulations suggest that dynamic climate control and metabolic heat and moisture reuse can improve the energy intensity of BSFL bioconversion. The quantitative estimates and hardware recommendations require experimental validation before application to industrial facilities.

## Conflict of Interest

The authors declare that they have no known competing financial interests or personal relationships that could have appeared to influence the work reported in this paper. This includes Moritz Gold who worked at REPLOID Group AG in parallel to ETH Zurich during execution of the presented research.

[a] ORCID: https://orcid.org/0009-0007-6113-0983
[b] ORCID: https://orcid.org/0000-0002-5975-5147
[c] ORCID: https://orcid.org/0000-0001-7285-6164
[d] ORCID: https://orcid.org/0000-0002-3398-1226

**Funding**

This research is financially supported by funds of the Federal Ministry of Agriculture, Food and Regional Identity (BMLEH) approved by the Parliament of the Federal Republic of Germany via the Federal Office for Agriculture and Food within the project 'Entwicklung nachhaltiger und wettbewerbsfähiger Insektenlebensmittel' OptiFood grant number 281A808D21. The publication of this article was funded by Chemnitz University of Technology.

**Acknowledgements**

The authors gratefully acknowledge the computing time made available to them on the high-performance computer at the NHR Center of TU Dresden. This centre is jointly supported by the Federal Ministry of Research, Technology and Space of Germany and the state governments participating in the NHR (www.nhr-verein.de/unsere-partner). We thank Alexander Kobelski for proofreading and his help in improving the readability of this work. During the preparation of this work, the authors used Claude (Anthropic) to improve the language, grammar and readability of the text. After using this tool, the authors reviewed and edited the content as needed and take full responsibility for the content of the publication.

**Data sharing**

Upon publication, data and code will be made public via GitHub.

**Author contributions**

**Sriram Saravanan:** Conceptualisation, Methodology, Software, Writing – original draft, Visualisation, Validation, Formal analysis, Investigation, Data curation, Resources. **Philipp Sauerteig**: Conceptualisation, Methodology, Writing – Review & Editing, Supervision, Investigation, Validation. **Moritz Gold:** Conceptualisation, Methodology, Writing – original draft, Writing – Review & Editing, Supervision, Investigation, Validation. **Stefan Streif:** Conceptualisation, Project administration, Funding acquisition, Supervision, Writing – Review & Editing, Validation. All authors have read and approved the final manuscript.

## 1. Introduction

Bioconversion using black soldier fly larvae (BSFL, *Hermetia illucens*) is an emerging approach for transforming nutrients from diverse organic waste and residues into valuable raw materials for feed, fertiliser, and various technical applications (Gold et al., 2018). Laboratory-scale BSFL growth models have highlighted the importance of controlling air temperature and humidity, and gas concentrations for efficient bioconversion and larval growth (Gligorescu et al., 2019; Padmanabha et al., 2020; Sripontan et al., 2020). This is because BSFL are ectothermic insects and their body temperature is largely influenced by the surrounding environment. In addition, water and gaseous metabolic by-products (primarily $CO_2$) need to be managed (Parodi et al., 2020; Zorrilla et al., 2026). Managing water and $CO_2$ is needed to allow evaporative cooling and keep the substrate sufficiently moist for digestibility. It also avoids process interruptions from larval escape caused by excessive $CO_2$ and air humidity. In addition, it keeps the substrate dry enough for larvae and frass to be sieved effectively at harvest (Fuhrmann et al., 2025). In efficient industrial and high-density larval bioconversion facilities, air temperature, humidity, and gas concentration are controlled using climate-system components such as ventilator, heater and cooler, and humidifier and dehumidifier (Bühler Group, 2025; Lalander et al., 2020).

During bioconversion, larvae generate substantial heat, and substrate temperatures can rise about 10 °C above the surrounding air (Zorrilla et al., 2026). This can be a challenge as high heat may reduce bioconversion efficiency due to increased metabolic costs, elevated microbial activity, higher mortality rates, and rapid drying of the substrate (Fuhrmann et al., 2025). Yet, this energy, a metabolic by-product of larval and microbial respiration together with frictional heat from their movement (Zorrilla et al., 2026), could also be used to meet energy requirements. The patterns of this heat production are still poorly understood, but key factors are known to include substrate nutrient content, larval density, biomass amount, and the substrate's physical properties (Klammsteiner et al., 2025; McEachern, 2018). Even though metabolic heat production per unit body mass decreases as BSFL advance through their larval stages (Gligorescu et al., 2019), total metabolic heat production per larva and per tray increases as larvae grow (Laganaro et al., 2021; McEachern, 2018). Consequently, low-energy-producing and high-energy-producing bioconversion stages could be combined in one container, or containers could be coupled, sharing resources between them.

Climate control is therefore a major cost, whether the system adds heat, removes it, or reuses the heat the larvae produce. In commercial-scale operations, heating, cooling, humidification, dehumidification and ventilation for larvae can account for 30-60% of the total energy expenditure (Bühler Group, 2025). Therefore, optimising these processes is a research priority to reduce operational costs and enhance the competitiveness of bioconversion products. One promising route is model-based control. The idea is to build a mathematical model of how the container behaves, that is, how temperature, humidity, and larval growth respond to the heater, ventilator, humidifier, and the other actuators. The controller uses this model to predict how the system will respond to a given set of actuator commands. It then picks the commands that best meet a chosen goal, such as using the least energy or growing the most biomass. This approach has already been tested for BSFL bioconversion at laboratory scale. Padmanabha et al. (2023) solved an optimal control problem (OCP) to improve larval growth while lowering energy use. Their controller ran open-loop: it computed actuator values once and applied them without checking the result against measurements. Such a controller works well only as long as the model resembles the real system it represents. Such deviations are common: the outdoor temperature may differ from the forecast, or the larvae may grow faster than predicted. Model Predictive Control (MPC) addresses this issue (Rawlings et al., 2017). At each step, MPC uses the model to look a fixed period ahead, called the prediction horizon, and computes the sequence of actuator values that best meets the goal over that period. Only the actuator values computed for the first step are applied to the real system; then the horizon shifts forward by one step and the whole calculation is repeated with the latest measurements received from the real system. Because the controller recomputes at every step using fresh information, it continuously adapts to disturbances and model error instead of following a fixed plan in advance. Al Khatib et al. (2022) developed an MPC to control the BSFL bioconversion. However, solving the OCP at every step was computationally demanding, particularly when the controller had to plan several days ahead while also reacting to fast outdoor climate changes. This computation cost grew with the size of the system. Sathyanarayanan et al. (2024) addressed this issue with a hierarchical control framework that split the problem into two levels: an upper level handled the long term planning, and a lower level tracked that plan while reacting to short-term changes. The authors applied the hierarchical control framework to control greenhouse. This kept the feedback benefits of MPC at a much lower computational cost.

These control approaches, however, have so far been demonstrated only at laboratory scale, and they have not been shown to carry over to large-scale industrial farms. Recent work on BSFL process optimisation has mainly addressed bioconversion in controlled laboratory reactors, small climate chambers and pilot facilities. Dynamic, control-oriented models for large-scale industrial farms and their climate systems remain very limited (Fuhrmann et al., 2025; Yakti et al., 2022). One promising industrial setup for larval bioconversion uses one or multiple shipping containers or rooms (20-40 foot). Shipping containers are globally available at low cost, come in standard sizes, and install quickly. They are also modular, capacity can be scaled simply by adding containers, and, with an adequate climate system, can provide a suitable bioconversion environment independent of the outdoor climate.

In this paper, we developed a control-oriented dynamic model of container-based BSFL bioconversion and used closed-loop simulations to investigate four operating and design questions. The studies used different subsets of eight climate scenarios:

1. How much does dynamic setpoint control improve energy use and biomass yield compared with fixed setpoints?
2. Which heating, ventilation and air conditioning (HVAC) component does a container need in a given climate, and are active cooling and dehumidification required?
3. Does operating two larval ages together in one container save energy, and which age difference works best?
4. Can this reuse be improved by placing the two ages in separate containers linked by an actuated air duct, so that heat transfer becomes controlled rather than passive?

## 2. Materials and methods

*Framework overview*

This study adopted a hierarchical control framework previously applied to two-timescale problems in controlled-environment agriculture (Daniels et al., 2024; Mansour et al., 2025; Sathyanarayanan et al., 2024). Industrial BSFL bioconversion is a two-timescale problem because it happens at a slow biological timescale, on which the larvae grow, and a fast climate timescale in which the HVAC responds to actuation. The lower level used a detailed model that captured larval growth and container climate on both timescales. The upper level used a simplified model that abstracted the fast climate dynamics, so that multi-day planning stayed tractable. Together, the detailed and simplified models formed the prediction model, which the controller used to plan and track. The detailed and simplified models are described in Supplementary material.

Because no physical container was available, all evaluations were performed in closed-loop simulation. An instance of the detailed model represented the virtual container, referred to here as the Digital Twin. This virtual container was not calibrated or validated against measurements from an industrial container farm.

*Modelling container-based BSFL bioconversion*

Figure 1 shows the BSFL bioconversion system housed in a 20-foot shipping container, divided into left (l) and right (r) zones along the airflow path. Each tray measured 0.6 m × 0.4 m and initially contained 15,000 larvae, 10 kg of water and 3 kg of substrate. Trays were arranged with their 0.4 m dimension across the container width and their 0.6 m dimension along its length. This allowed five tray positions across the width and nine along the length, with 10 horizontal layers. Three of the five positions across the width belonged to the left zone and two to the right, giving 270 trays on the left (3 × 9 × 10) and 180 on the right (2 × 9 × 10). The unequal tray numbers resulted from using all available positions in this layout. For modelling, all trays within each zone were combined into one equivalent tray, preserving their total surface area, volume, substrate mass and larval count. Each zone was treated as well mixed, with larvae of the same age following the same growth trajectory.

The container climate was controlled by an air recirculation loop. Two perforated walls were placed along the long sides of the container. The left wall served as the air inlet and the right wall as the outlet. Air left through the right wall, passed through an external recirculation pipe connecting the two long sides, and re-entered through the left wall. The HVAC components (heater/cooler, ventilator, humidifier, and dehumidifier) were mounted on the recirculation pipe and conditioned the air before it re-entered the container. A door on one short side allowed access for harvesting and for starting a new bioconversion cycle.

Conditions inside the container were not spatially uniform. Air flowed from left to right and picked up heat, moisture and $CO_2$ from the substrate and larvae along the way. The model captured this by treating each zone as its own well-mixed subsystem, coupled to its neighbouring zone and to the recirculation pipe via the airflow.

An industrial BSFL bioconversion system typically consists of multiple containers operated side by side, each at a different stage of its bioconversion cycle. This creates an opportunity for energy reuse: surplus heat and humidity from one container can be routed to another to reduce the total HVAC load. We modelled the simplest coupling technique, two containers connected by a single duct with a controllable coupling actuator between their recirculation pipes (Figure 1). Extension to larger networks would require further modelling and evaluation.

The biological growth model was adapted from Padmanabha et al. (2020), who validated it against small-scale laboratory data. Parameters were scaled to represent the container configuration considered here; this scaling does not constitute validation at industrial scale. The model derivations and parameter values are provided in the supplementary material.

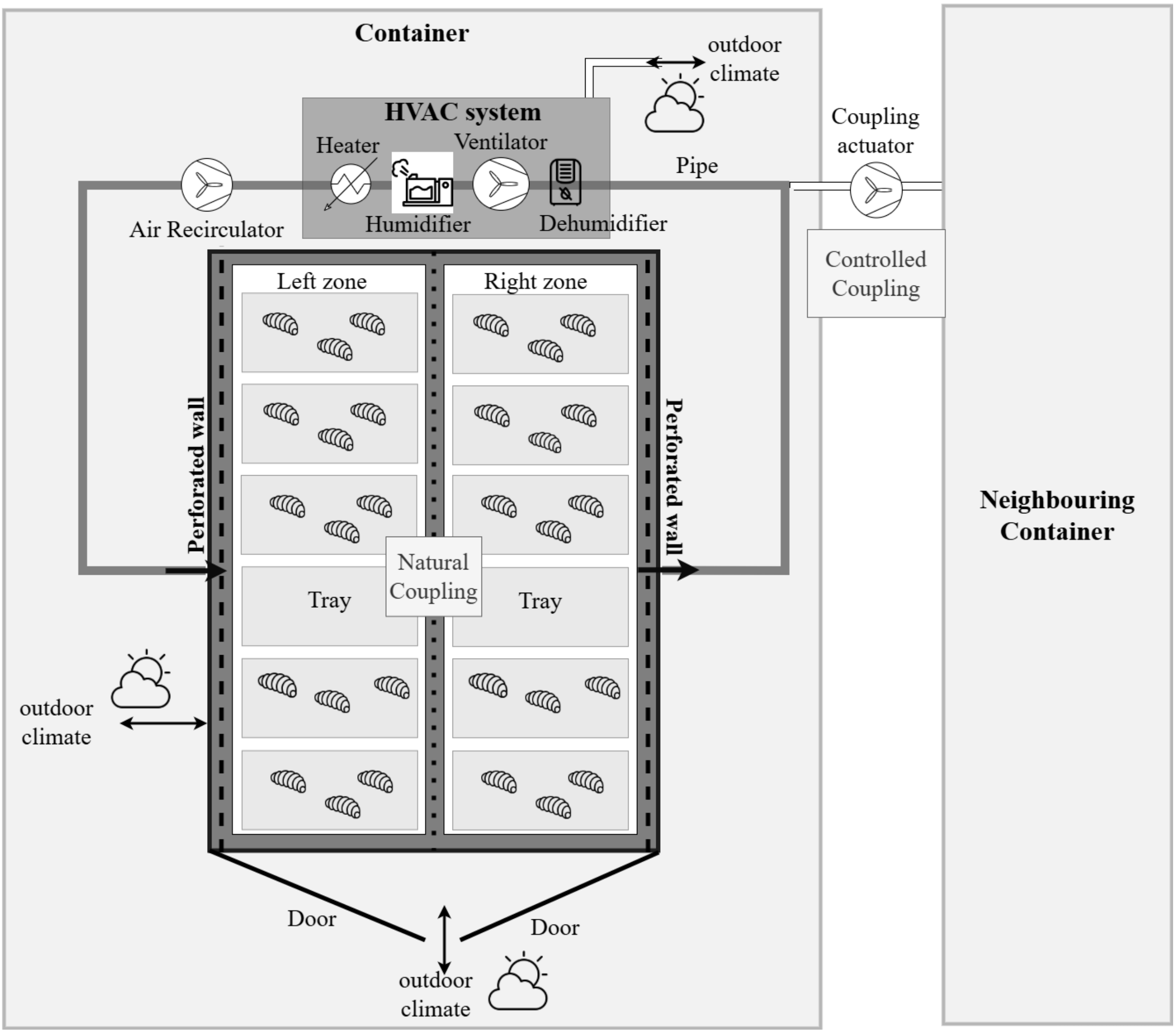

**Figure** *1*. Container model. Subsystems of the container model and their interactions, shown in top view. Two zones, left and right, are separated by the dotted centre line, and each holds multiple trays containing substrate and larvae. The Heating, Ventilation and Air Conditioning (HVAC) system conditions the air, and the recirculation unit drives it through the pipe. Air enters through the left perforated side wall and leaves through the right one (dashed lines; arrows show the flow direction). The container exchanges heat, $CO_2$ and moisture with the outdoor climate. The two zones share the same air, so heat, $CO_2$ and moisture pass passively between them (double-headed arrow); we refer to this as natural coupling. A neighbouring container can be connected through the actuated air duct, so that heat, $CO_2$ and moisture are transferred between two containers; we refer to this as controlled coupling.

*Hierarchical control*

Figure 2 presents the hierarchical controller with two levels. The upper level handled long-term planning. Once per day, it determined the hourly climate setpoints to maintain for the next 7 days and decided when to harvest. The harvest time was chosen from three candidates: the end of day seven, or 8 hours or 16 hours before that. Since the upper level looked several days ahead, using the detailed model here would have been computationally very costly; the upper level therefore used the simplified model. The lower level handled short-term regulation. It adjusted the HVAC actuators multiple times per hour to track upper-level setpoints of temperature and humidity in real time amid outdoor climate acting as a disturbance on the system. For this task, the detailed model was used, which also captured the fast actuator dynamics.

The two levels followed a receding-horizon approach but used different models, objectives and timescales. The upper-level problem was solved once per day, producing hourly setpoints over a 7-day prediction horizon. The lower-level problem was solved every 15 minutes over a 1 hour prediction horizon. At each update, the controller optimised its control sequence subject to state and input bounds, using the latest information from the Digital Twin. The lower level applied the first actuator command and then repeated the calculation at its next update. The upper level replanned the climate trajectories for the next 7 days in a receding fashion and reconsidered the harvest time as new state information became available.

The upper level planned temperature and humidity trajectories by balancing predicted dry biomass at harvest, residual substrate water mass and control effort. The lower level balanced tracking of these trajectories against actuator effort. Neither level optimised electrical energy directly. Penalising HVAC usage limited energy use indirectly, and the energy reported in this paper was calculated after simulation from the rated power of each actuator. For two coupled containers, both levels treated the pair as a joint system: the upper level optimised the climate trajectories and coupling strength, and the lower level tracked the trajectories and coupling command. The detailed optimisation formulations are provided in the supplementary material.

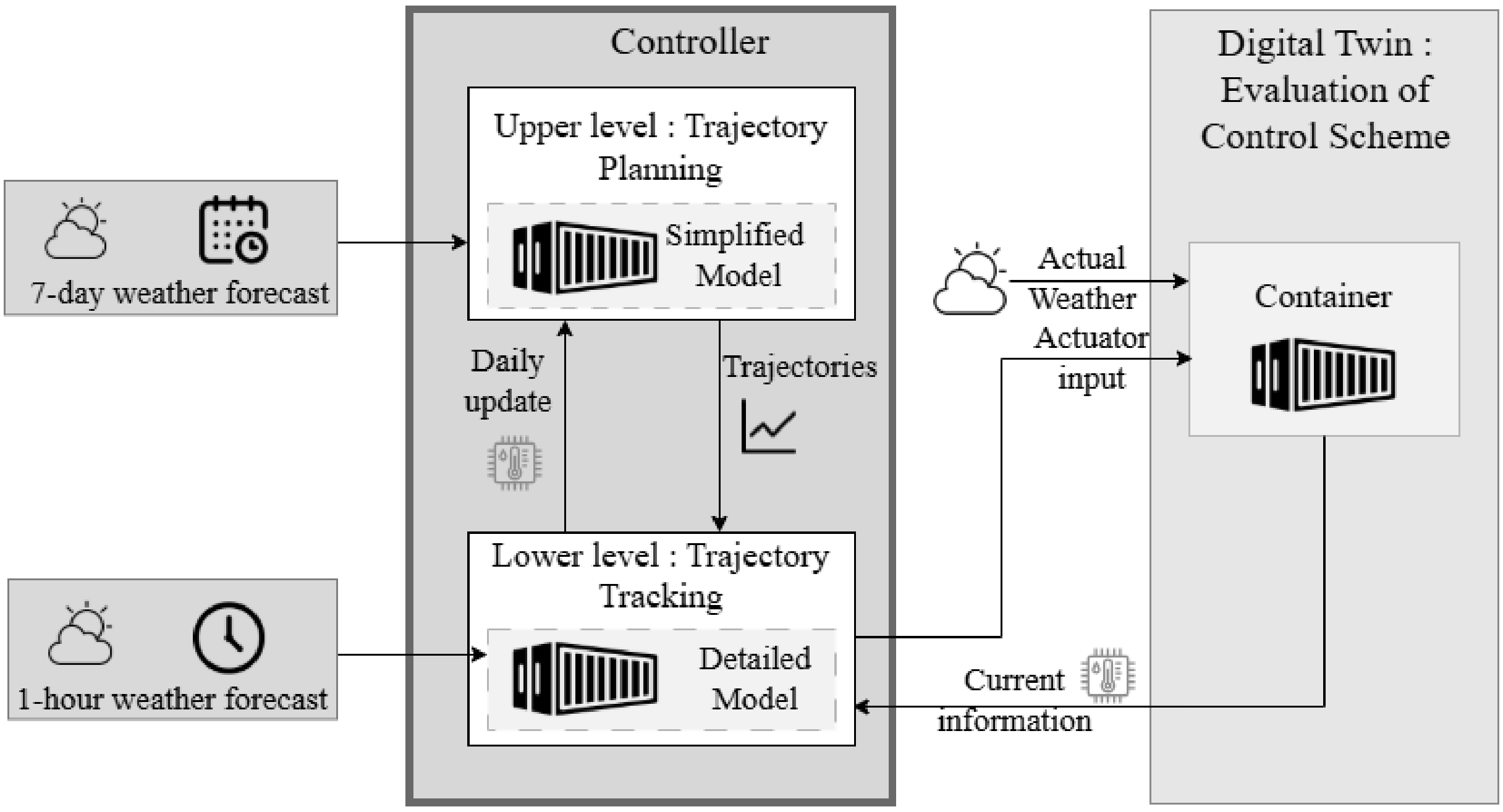

**Figure 2**. Hierarchical Control Framework. Hierarchical control framework for container-based BSFL bioconversion. The framework has two levels that operate at different timescales and exchange information in a closed loop. Once per day, the upper level uses the simplified model to compute hourly temperature and humidity trajectories for the next 7 days. It passes these trajectories to the lower level, which uses the detailed model to compute the HVAC actuator values every 15 minutes and applies them to the Digital Twin. The Digital Twin stands in for the real container in the closed-loop simulation. It returns its current information to the lower level, closing the inner loop, and the lower level passes a daily update to the upper level, closing the outer loop. Outdoor weather acts on both the controller and the Digital Twin.

The hierarchical controller operated as follows (Figure 2): once per day, the upper level received the current Digital Twin information and solved the upper-level OCP for each candidate harvest time, reselecting the harvest time. Because this was repeated daily with updated information, the harvest time could change from day to day when growth deviated from the earlier prediction. The resulting setpoint trajectory and harvest time were passed to the lower level. For simplicity, we assumed that the necessary information from the Digital Twin, such as temperature and biomass, was available.

*Model input parameters across all experiments*

The prediction models used the parameter set of Padmanabha et al. (2020), with the container-scale adaptations listed in the supplementary material. To introduce a limited mismatch between the prediction and simulation models, two biological parameters were changed in the Digital Twin: the substrate moisture above which larvae cease to grow, and the maximum growth rate of a larva. This perturbation tested the controller under one specified model mismatch; it did not represent the full uncertainty associated with feed composition, larval population or scale-dependent heat and mass transfer. The parameter definitions, nominal values and altered values are listed in the supplementary material.

The controller was evaluated across the climates an industrial bioconversion facility might face, from temperate Northern European deployments to hot equatorial ones. To cover these conditions, we defined eight scenarios spanning four temperature ranges and two humidity variants (Table 1). The dry variants corresponded to outdoor relative humidity between 10% and 40%, and the wet variants to between 70% and 95%. For each scenario, we used a four-week segment of historic climate data from a public dataset (Inness et al., 2019), sampled at 15 minutes intervals. The stated temperature and humidity values were the minimum and maximum values observed over that segment.

**Table 1**. List of climate scenarios with their temperature, and humidity ranges and country where they occur.

| **Climate scenario** | **Temperature range** | **Relative humidity range** | **City, Country** | **Month** |
|---|---|---|---|---|
| $Cool_{dry}$ | 4-10 °C | 10-40% | Yellowknife, Canada | December - January |
| $Mild_{dry}$ | 14-19 °C | 10-40% | Madrid, Spain | May and October |
| $Warm_{dry}$ | 24-28 °C | 10-40% | Cairo, Egypt | April and November |
| $Hot_{dry}$ | 35-38 °C | 10-40% | Riyadh, Saudi Arabia | June - September |
| $Cool_{wet}$ | 4-10 °C | 70-95% | London, United Kingdom | November - March |
| $Mild_{wet}$ | 14-19 °C | 70-95% | Auckland, New Zealand | April and November |
| $Warm_{wet}$ | 24-28 °C | 70-95% | Singapore, Singapore | Year round |
| $Hot_{wet}$ | 35-38 °C | 70-95% | Dubai, United Arab Emirates | July - September |

The two levels in the hierarchical controller did not have access to the actual climate data listed in Table 1. Each level operated on a weather forecast, generated by adding noise to the actual climate data. The upper level received a 7-day weather forecast and the lower level a 1 hour weather forecast. The forecast noise magnitudes are given in supplementary material.

*Experimental design*

Four simulation experiments addressed controller selection (E1), HVAC configuration (E2), larval age pairing (E3) and inter-container coupling (E4). They were performed sequentially, and selections from earlier experiments were used in subsequent experiments. The controller selected in E1 was used in E2–E4, and the climate-specific HVAC configurations selected in E2 were used in E3.

*Experiment 1 (E1) – Static setpoint vs. dynamic-setpoint control*

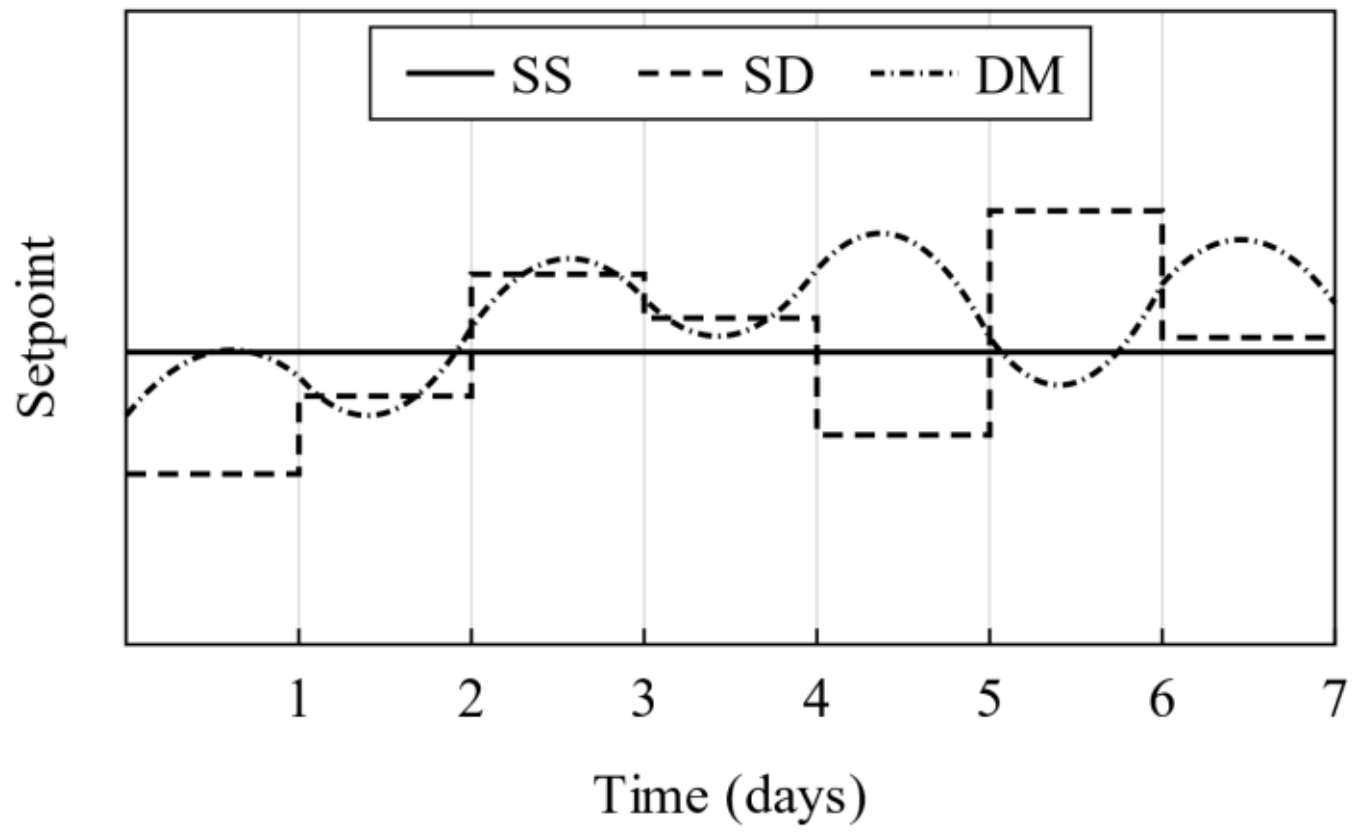

**Figure 3**. Setpoint behaviour of the three controllers. Qualitative behaviour of the three controllers compared in the Experiment 1 (E1), shown over one bioconversion cycle. Single Setpoint (SS) holds a single setpoint for the entire bioconversion cycle; Setpoint Daily (SD) updates the setpoint once per day and holds it constant over each 24-hour window; Dynamic MPC (DM) varies the setpoint continuously through the cycle.

The experiment was designed to determine which controller was best suited for container-based BSFL bioconversion. The experiment compared a static setpoint strategy, which, to the authors' knowledge, is common in current industrial practice, against a dynamic setpoint strategy.

Three controllers were compared, differing only in how the upper level set the climate setpoints over the bioconversion cycle (Figure 3):

- Single Setpoint (SS): the upper-level OCP was solved once at the start of the bioconversion cycle. The result was a single temperature and humidity setpoint, which was then held constant for the whole bioconversion cycle.
- Setpoint Daily (SD): the same OCP as in SS was solved, but at the start of each day using the latest Digital Twin information. The resulting setpoint pair was held constant for that day, and could differ from one day to the next depending on the Digital Twin information.
- Dynamic MPC (DM): the upper-level OCP was solved at the start of each day, as in SD, but the solution was not restricted to a single setpoint. It was a full trajectory over the horizon, so temperature and humidity could vary at every step.

All three controllers shared the same lower level and detailed model.

The controllers were compared under four climate scenarios: $\text{Cool}_{\text{dry}}$, $\text{Mild}_{\text{wet}}$, $\text{Warm}_{\text{dry}}$ and $\text{Hot}_{\text{wet}}$. This subset covered all four temperature ranges and both humidity variants, but did not constitute a complete factorial comparison across all eight scenarios. The controller selected from these four comparisons was used in subsequent experiments. Its ranking in the remaining four scenarios was not independently tested.

Both zones grew larvae of identical age and on identical substrate, so any performance difference was attributable to the controller. Each scenario was simulated for one bioconversion cycle lasting 7 days. A bare minimum HVAC configuration was used, consisting of a heater, a ventilator, an air recirculation unit and a humidifier.

*Experiment 2 (E2) – HVAC configuration selection*

In E2, four HVAC configurations were compared across all eight climate scenarios (Table 1), using the DM controller selected in E1. The configurations were:

- Baseline: heater, humidifier, ventilator and air recirculation unit, without active cooling or dehumidification.
- +Cooler: adds active cooling,
- +Dehumidifier: adds active dehumidification,
- +Cooler+Dehumidifier: adds both.

As in E1, both zones in the container grew larvae of identical age and on identical substrate, isolating the effect of HVAC configuration from inter-zone heterogeneity. Each scenario was simulated for one bioconversion cycle lasting 7 days.

*Experiment 3 (E3) – Age-pair selection*

In this experiment, we examined how BSFL bioconversion of two different ages in the two zones of a single container affected energy consumption, biomass yield, and frass water at harvest. We also tested how the energy consumption depended on the age pair and the climate scenario. Because the two larval ages shared one container, surplus metabolic heat and moisture from one age could passively offset the other's demand. $CO_2$ produced in either zone was also shared between them. We refer to this within-container sharing as *natural coupling*, as there is no barrier between the zones (see Figure 1).

We used four of the eight climate scenarios: $\text{Cool}_{\text{dry}}$, $\text{Mild}_{\text{wet}}$, $\text{Warm}_{\text{dry}}$, and $\text{Hot}_{\text{wet}}$. Each simulation was run with the HVAC configuration identified as necessary for its climate in E2.

Candidate age pairs (a, b), where a and b denote the age in the left and right zone of the container respectively, were compared against a reference age pair (1, 1):

$$\text{Age pairs} = \{(1,2), (1,3), (1,4), (1,5), (1,6), (1,7)\}.$$

Age denotes the number of days since the start of the bioconversion cycle, rather than age since hatching. Before harvesting or restarting either zone, an initial pair (1,5) therefore progresses to (2,6) one day later. For each age pair (1, b), the right zone was initialised with the state of a simulation of the reference pair (1,1) at day b, including its larval biomass and remaining substrate. Each simulation lasted 7 days. Each zone was harvested at the end of its own 7-day cycle and immediately restarted with fresh larvae, so every zone was harvested exactly once within the simulation window. Biomass was taken as the total dry biomass harvested from both zones, and HVAC energy was counted over the full window.

Two tray-loading configurations were considered. The asymmetric case used all 450 tray positions in the selected layout: 270 in the left zone and 180 in the right zone. This corresponded to loading unequal batches on the two loading days. In the symmetric case, the left zone held 18 rather than 27 trays per horizontal layer, giving 180 trays in each zone and 360 in total. Some tray positions therefore remained unused, while the container air volume was unchanged. Larval ages, substrate composition, initial loading per tray and HVAC configuration remained the same, but the total number of larvae and the initial substrate and water quantities were reduced by 20%. The symmetric case served two purposes. First, it tested whether the effect of age pairing depended on the unequal zone sizes. Second, it showed the energy cost of partial loading: the HVAC system still conditions the same air volume, while fewer larvae contributed biomass and metabolic heat.

For any scenario in E2, where HVAC configuration other than the Baseline was selected, a component trade-off check was performed. The best age pair for that scenario was also simulated with Baseline HVAC configuration, to test whether the metabolic heat reuse from age-pairing could replace the added component.

*Experiment 4 (E4) – Inter-container coupling*

E4 compared two facility arrangements: mixed-age zones within two independent containers, and age-separated containers connected by an actuated air duct. The latter arrangement allowed deliberate air exchange and whole-container harvesting. The comparison therefore evaluated the combined effects of coupling, age allocation and harvesting strategy, rather than the isolated effect of the duct. Energy use and biomass were assessed; investment costs, maintenance costs and financial payback were not quantified.

Two coupling techniques were compared:

- **Natural coupling:** two containers operated independently. Each container held the best asymmetric-case age pair from E3 in its two zones. Within a container, the two ages exchanged heat and moisture passively through the shared air, but the two containers were not connected to each other.
- **Controlled coupling:** two connected containers, one initially holding young larvae in both zones and the other initially holding older larvae in both zones. The controller regulated air transfer through the actuated duct (Figure 1).

The experiment used only the $Cool_{dry}$ scenario, where heating was the dominant load and heat reuse between containers was therefore most relevant. The experiment was simulated for 30 days, covering roughly four 7-day bioconversion cycles per container.

*Performance metrics*

The following performance metrics were used across all experiments. The HVAC energy consumption $E_{HVAC}$ is the cumulative electrical energy drawn by the heater/cooler, ventilator, humidifier, dehumidifier, and air recirculation unit over the bioconversion cycle or the entire simulation and is calculated as:

$$E_{HVAC} = \sum_{k=1}^{N_{sim}} \left( P_T|u_T(k)| + P_V u_V(k) + P_D u_D(k) + P_A|u_A(k)| + P_c u_C(k)\right)\Delta t + E_{hum} + E_{dehum} \quad (1)$$

Where $N_{sim}$ is the total simulation time, $\Delta t$ is the sampling time; $P_T, P_V, P_H, P_D, P_A$ and $P_c$ are the rated electrical power of the heater/cooler, ventilator, humidifier, dehumidifier, air recirculation unit and coupling actuator respectively; and $u_T, u_V, u_H, u_D, u_A, u_C$ are their normalised control signals. The humidifier and dehumidifier energies, $E_{hum}$ and $E_{dehum}$, were calculated from the latent heat of the water vapour each unit adds or removes, divided by its efficiency (supplementary material). $E_{HVAC}$ is measured in kWh.

The specific energy consumption (SEC, kWh/kg) is the HVAC energy used within the simulation window divided by the harvested dry biomass in that window:

$$SEC = \frac{E_{HVAC}}{B_{harvest}}$$

where $B_{harvest}$ is the sum over both zones and all harvest events in the window of the number of larvae multiplied by the predicted dry mass per larva at harvest. The number of larvae was fixed, so no separate survival factor was applied. Losses during separation of larvae from frass were not modelled, so $B_{harvest}$ is an upper bound on recoverable biomass. SEC therefore expresses the modelled electrical energy consumption per kilogram of predicted dry larval biomass within the defined HVAC system boundary (Lawrence et al., 2019). It does not represent the total energy demand of the full production and processing chain.

The frass water is the moisture of the substrate at harvest, expressed as a percentage:

$$\text{Frass water (\%)} = \frac{W_{med}^{harvest}}{W_{med}^{harvest} + N_{med}^{harvest} + N_{exc}^{harvest}} \times 100\%$$

where $W_{med}^{harvest}$ is the frass water mass at harvest, $N_{med}^{harvest}$is the nutrients mass available at harvest and $N_{exc}^{harvest}$ is the larval excretion mass at harvest. Frass with a water content below 50% is treated as sievable and above 50% not sievable, following the separation threshold reported for BSFL frass (Cheng et al., 2017).

*Software and data analysis*

Simulations were implemented in MATLAB 2024a (The MathWorks Inc., 2024) using the CasADi optimisation toolbox (Andersson et al., 2019). The optimisation problems were solved with IPOPT (Biegler and Zavala, 2009) using the Harwell Subroutine Library's (HSL) linear solver MA57 (Duff, 2004).

## 3. Results and discussion

*E1 – Static setpoint vs. dynamic setpoint control*

In E1, we compared the three controllers (SS, SD, DM) across four climate scenarios. For each scenario, biomass, SEC, and frass water are reported in Table 2.

DM achieved the lowest SEC in all four tested climate scenarios and the highest predicted dry biomass except in $Mild_{wet}$ (Table 2). Relative to SS, biomass increased by 27.5% in $Cool_{dry}$, 22.8% in $Mild_{wet}$, 24.8% in $Hot_{wet}$ and about 2% in $Warm_{dry}$. SEC decreased by 40%, 42%, 61% and 31%, respectively.

Outdoor conditions affected the container through wall heat transfer, door opening and ventilation. All three strategies used the same lower-level feedback controller and could therefore adjust HVAC actuation in response to disturbances. Their distinction was the flexibility of the upper-level climate reference: SS used one fixed setpoint pair, SD updated it daily, and DM planned hourly-varying trajectories. The results are consistent with an advantage from adapting these trajectories to forecast weather and predicted larval development. However, the relative contributions of weather adaptation, model mismatch and harvest-time selection were not isolated in E1.

Model-based control of insect bioconversion has so far been demonstrated only at laboratory scale. The closest industry-scale comparison therefore comes from other controlled-environment systems, such as greenhouses. In greenhouse climate control, dynamic setpoints have been shown to reduce energy use and increase yield compared with static setpoints (Chen et al., 2022; Su and Xu, 2025). Padmanabha et al. (2023) and Mansour et al. (2026) reported the same outcome for insect bioconversion, though at laboratory scale and without the energy and frass-water metrics used here. To our knowledge, no study has evaluated dynamic climate control against static setpoints at container or industrial scale using these metrics, which is the gap E1 addresses.

**Table 2**. Biomass, Specific Energy Consumption (SEC), and frass water at harvest for the three controllers (SS – Single Setpoint; SD – Setpoint Daily; DM – Dynamic MPC) across the four climate scenarios (E1). Bold values indicate more than 50% of frass water.

| Metric | Controller | $\mathbf{Cool_{dry}}$ | $\mathbf{Mild_{wet}}$ | $\mathbf{Warm_{dry}}$ | $\mathbf{Hot_{wet}}$ |
|---|---|---|---|---|---|
| **Biomass (kg)** | SS | 220.0 | 220.0 | 267.0 | 125.0 |
| | SD | 263.0 | 271.0 | 269.4 | 149.0 |
| | DM | 280.4 | 270.2 | 272.3 | 156.0 |
| **SEC (kWh/kg)** | SS | 10.0 | 9.8 | 10.1 | 40.3 |
| | SD | 9.8 | 11.8 | 9.1 | 36.3 |
| | DM | 6.0 | 5.7 | 7.0 | 15.9 |
| **Frass water (%)** | SS | 49.5 | 48.2 | 49.0 | **64.0** |
| | SD | 30.0 | 44.0 | 21.0 | **58.0** |
| | DM | 33.2 | 23.7 | 22.6 | **64.2** |

SD did not consistently achieve an SEC between those of SS and DM. In $Mild_{wet}$, SD produced slightly more biomass than DM (271.0 versus 270.2 kg), but used substantially more energy per kilogram (11.8 versus 5.7 kWh/kg). Its SEC also exceeded that of SS (9.8 kWh/kg).

All three controllers met the assumed frass-water criterion for sieving (< 50%) in $Cool_{dry}$, $Mild_{wet}$ and $Warm_{dry}$. None met it in $Hot_{wet}$ (Table 2). This classification is based on the modelled water content and the adopted threshold, not on measured separation performance.

The $Hot_{wet}$ scenario was an outlier in every metric. Its biomass yield was roughly half that of the other scenarios and its SEC was two to four times higher than the other scenarios. This was because the HVAC had no active cooling in this experiment. At outdoor temperatures of 35–38 °C, no controller could therefore bring the container into the larval growth optimum (Padmanabha et al., 2020). Without active cooling, the metabolic heat released by the larvae raised the container temperature above the outdoor temperature. Because warmer air can hold more water vapour, the relative humidity in the container decreased even though the outdoor air was humid. The humidifier therefore had to add water to keep the humidity at its setpoint, and at these high temperatures each unit of humidifier input evaporated more water. Humidification therefore accounted for most of the energy use in this scenario, which explains its high SEC. Nevertheless, DM performed better than the other two controllers in this scenario.

DM was selected for E2–E4 because it achieved the lowest SEC in all four tested scenarios, not because it was superior in every metric. E1 also indicated that controller choice alone could not resolve the low biomass and high frass-water content observed in $Hot_{wet}$ without active cooling. This motivated the HVAC comparison in E2.

*E2 – HVAC configuration selection*

In E2, we compared four HVAC configurations across eight climate scenarios, using DM and the single-container, same-age setting from E1. Biomass, SEC, and frass water at harvest are reported in Table 3.

**Table 3**. (a) Biomass (kg), (b) Specific Energy Consumption (SEC) (kWh/kg), and (c) Frass water, for the four HVAC configurations across the eight climate scenarios.

| Scenario\ Configuration | Baseline | +Cooler | +Dehumidifier | +Cooler +Dehumidifier |
|---|---|---|---|---|
| **(a) Biomass (kg)** | | | | |
| $Cool_{wet}$ | 289.7 | 289.7 | 294.9 | 286.2 |
| $Cool_{dry}$ | 280.4 | 277.3 | 269.5 | 270.3 |
| $Mild_{wet}$ | 270.2 | 265.9 | 269.2 | 282.4 |
| $Mild_{dry}$ | 277.5 | 280.0 | 264.3 | 279.0 |
| $Warm_{wet}$ | 272.3 | 283.0 | 266.3 | 274.5 |
| $Warm_{dry}$ | 272.3 | 269.4 | 266.2 | 281.3 |
| $Hot_{wet}$ | 156.0 | 285.2 | 167.7 | 273.1 |
| $Hot_{dry}$ | 169.3 | 261.1 | 163.4 | 258.1 |
| **(b) SEC (kWh/kg)** | | | | |
| $Cool_{wet}$ | 5.9 | 5.6 | 9.0 | 8.7 |
| $Cool_{dry}$ | 6.0 | 6.3 | 9.7 | 9.7 |
| $Mild_{wet}$ | 5.7 | 5.3 | 10.9 | 11.1 |
| $Mild_{dry}$ | 7.7 | 6.8 | 9.3 | 9.0 |
| $Warm_{wet}$ | 4.7 | 5.4 | 15.9 | 21.0 |
| $Warm_{dry}$ | 7.0 | 7.4 | 10.4 | 11.3 |
| $Hot_{wet}$ | 15.9 | 9.8 | 18.0 | 17.6 |
| $Hot_{dry}$ | 20.5 | 6.7 | 24.0 | 16.5 |
| **(c) Frass water (%)** | | | | |
| $Cool_{wet}$ | 44.2 | 44.1 | 46.4 | 42.0 |
| $Cool_{dry}$ | 33.2 | 42.6 | 38.6 | 40.7 |
| $Mild_{wet}$ | 23.7 | 22.7 | 30.0 | 37.4 |
| $Mild_{dry}$ | 38.9 | 39.6 | 32.2 | 41.7 |
| $Warm_{wet}$ | 27.1 | 29.3 | 20.5 | 23.9 |
| $Warm_{dry}$ | 22.6 | 24.2 | 24.9 | 35.4 |
| $Hot_{wet}$ | 64.2 | 35.9 | 45.9 | 23.5 |
| $Hot_{dry}$ | 36.7 | 22.8 | 32.1 | 31.8 |

In the six non-hot scenarios, all HVAC configurations produced biomass within 10% of the Baseline value (Table 3a). Frass-water content varied between configurations but remained below the assumed 50% sieving threshold in every case (Table 3c). Additional cooling or dehumidification therefore did not have a large impact on harvested larval biomass in these scenarios.

SEC differed more strongly between configurations. Under the tested control settings, both configurations with a dehumidifier had higher SEC than the Baseline in all six non-hot scenarios (Table 3b). Baseline and +Cooler differed by approximately 5–7% in $Cool_{wet}$, $Cool_{dry}$, $Mild_{wet}$ and $Warm_{dry}$. The differences were larger in $Mild_{dry}$, where +Cooler had about 12% lower SEC, and $Warm_{wet}$, where it had about 15% higher SEC. These comparisons indicate that Baseline was operationally sufficient in the tested non-hot scenarios, but not that it always minimised SEC. The results apply to the combined hardware and controller configuration; they do not demonstrate that optional cooling or dehumidification is intrinsically detrimental.

$Hot_{wet}$ and $Hot_{dry}$ scenarios behaved differently. Their outdoor temperature range (35–38 °C) lay near and above the assumed larval developmental optimum of around 35 °C, beyond which the development rate of larvae falls sharply (Padmanabha et al., 2020, Supplementary Figure S1). The metabolic heat released by the larvae and microbiome increases the substrate heat (Zorrilla et al., 2026). Without active cooling the substrate overheated and biomass yield was between 156-169 kg for Baseline and +Dehum (Table 3a). +Cooler and +Cooler+Dehum instead held the substrate temperature near the optimum and increased the biomass yield to 258-285 kg. Cooling therefore both increased biomass and prevented overheating of the substrate. This resulted in SEC reduction relative to Baseline by 38% in $Hot_{wet}$ and 67% in $Hot_{dry}$ (Table 3b). Cooling also mattered for sievability, but only in $Hot_{wet}$, where +Cooler brought frass water from 64.2% to 35.9%, below the 50% limit (Table 3c). In the $Hot_{dry}$ scenario, the frass was already sievable in Baseline configuration at 36.7%. In $Hot_{wet}$, the dehumidifier in the +Cooler +Dehum configuration lowered the frass water even further to 23.5% , but frass was already sievable without it. So, this frass water reduction brought no practical benefit while increasing SEC. A dehumidifier was therefore not required in these climate scenarios.

Baseline was retained for the six non-hot scenarios because it maintained biomass and met the assumed frass-water criterion without additional active cooling or dehumidification. +Cooler was retained for $Hot_{wet}$ and $Hot_{dry}$ because cooling substantially increased predicted biomass; in $Hot_{wet}$ it also brought frass-water content below the assumed sieving threshold. Dehumidification was not retained under the tested control settings. These are scenario-specific selections, not a demonstrated economic optimum, because component costs and payback were not assessed. The selected configurations were used in E3.

*E3 – Age-pair selection*

E3 compared six mixed-age pairs with the same-age reference (1,1), in asymmetric and symmetric zone configurations, across four climate scenarios. Table 4 reports predicted dry biomass, SEC and frass-water content; Figure 4 shows the component energy consumption for the asymmetric configuration.

**Table 4**. (a) Biomass (b) Specific Energy Consumption (SEC) (kWh/kg) (c) Frass water (%) for each (scenario, case, age-pair) combination.

| **Scenarios\Age-Pair** | | **(1,1)** | **(1,2)** | **(1,3)** | **(1,4)** | **(1,5)** | **(1,6)** | **(1,7)** |
|---|---|---|---|---|---|---|---|---|
| **(a) Biomass (kg)** | | | | | | | | |
| **$Cool_{dry}$** | **asymmetric case** | 280.4 | 273.7 | 266.8 | 270.9 | 271.4 | 270.0 | 272.2 |
| | **symmetric case** | 216.7 | 220.1 | 214.4 | 218.6 | 218.6 | 219.4 | 215.4 |
| **$Mild_{wet}$** | **asymmetric case** | 270.2 | 262.0 | 270.0 | 271.2 | 272.9 | 268.6 | 272.6 |
| | **symmetric case** | 226.2 | 213.2 | 219.7 | 219.0 | 223.5 | 220.3 | 221.7 |
| **$Warm_{dry}$** | **asymmetric case** | 272.3 | 280.8 | 284.5 | 291.0 | 284.1 | 279.5 | 278.3 |
| | **symmetric case** | 210.2 | 219.6 | 229.0 | 230.2 | 227.1 | 228.7 | 230.5 |
| **Hot$_{wet}$** | **asymmetric case** | 285.2 | 284.9 | 292.9 | 295.8 | 297.6 | 293.8 | 290.8 |
| | **symmetric case** | 237.3 | 239.7 | 241.7 | 240.8 | 244.5 | 242.4 | 244.9 |
| **(b) SEC (kWh/kg)** | | | | | | | | |
| **Cool$_{dry}$** | **asymmetric case** | 6.0 | 5.5 | 5.5 | 4.9 | 4.9 | 5.4 | 5.5 |
| | **symmetric case** | 7.9 | 7.4 | 6.4 | 5.7 | 5.4 | 6.1 | 6.7 |
| **Mild$_{wet}$** | **asymmetric case** | 5.7 | 5.7 | 5.2 | 4.4 | 4.3 | 4.8 | 5.4 |
| | **symmetric case** | 6.9 | 6.7 | 5.5 | 4.9 | 4.8 | 5.3 | 6.2 |
| **Warm$_{dry}$** | **asymmetric case** | 7.0 | 6.0 | 5.4 | 4.8 | 4.7 | 5.2 | 6.6 |
| | **symmetric case** | 9.6 | 8.2 | 6.9 | 5.9 | 6.0 | 6.5 | 8.5 |
| **Hot$_{wet}$** | **asymmetric case** | 9.8 | 8.7 | 9.2 | 9.6 | 9.1 | 9.1 | 8.8 |
| | **symmetric case** | 11.0 | 10.0 | 10.5 | 10.3 | 10.4 | 10.3 | 9.7 |
| **(c) Frass water (%)** | | | | | | | | |
| **Cool$_{dry}$** | **asymmetric case** | 33.2 | 44.3 | 28.4 | 41.2 | 36.7 | 38.6 | 32.8 |
| | **symmetric case** | 45.0 | 33.5 | 28.6 | 39.6 | 40.7 | 39.5 | 39.3 |
| **Mild$_{wet}$** | **asymmetric case** | 23.7 | 25.5 | 29.7 | 35.9 | 32.3 | 29.3 | 33.2 |
| | **symmetric case** | 29.5 | 27.4 | 31.9 | 34.1 | 34.2 | 31.2 | 30.8 |
| **Warm$_{dry}$** | **asymmetric case** | 22.6 | 34.1 | 32.3 | 35.5 | 32.9 | 29.1 | 40.1 |
| | **symmetric case** | 18.0 | 35.6 | 35.8 | 35.8 | 34.6 | 32.5 | 32.7 |
| **Hot$_{wet}$** | **asymmetric case** | 35.9 | 36.3 | 34.6 | 36.7 | 38.0 | 40.7 | 35.2 |
| | **symmetric case** | 31.4 | 35.2 | 37.8 | 38.5 | 38.7 | 37.1 | 35.0 |

Predicted biomass across age pairs remained approximately within 94–110% of the corresponding same-age reference (Table 4a). Because the biomass changes were modest relative to the SEC differences and all treatments met the assumed frass-water criterion, SEC was used to compare candidate age pairs.

The asymmetric case achieved lower SEC than the symmetric case for every age pair and climate scenario (Table 4b). The two cases differed not only in how the trays were split between zones but also in total loading: 450 trays in the asymmetric case and 360 in the symmetric case. At pair (1,1), total HVAC energy was similar in both cases (within 7%), because the HVAC system conditioned the same air volume. Biomass, however, was 16–23% lower in the symmetric case (Table 4a). The higher SEC of the symmetric case was therefore mainly caused by the smaller harvested biomass, with possibly a smaller contribution from less internally generated heat and moisture. The comparison does not separate these effects and does not show an advantage of unequal zone sizes as such. Despite the higher SEC, the symmetric case showed the same pattern of energy reduction across age pairs as the asymmetric case.

In both cases, intermediate age offsets gave the lowest SEC in Cooldry, Mildwet and Warmdry (Table 4b). In the asymmetric case, pair (1,5) achieved 4.9, 4.3 and 4.7 kWh/kg, respectively, corresponding to reductions of about 18%, 25% and 33% relative to pair (1,1). In Cooldry, (1,4) and (1,5) were tied at 4.9 kWh/kg. In the symmetric case, the reductions were larger: (1,5) achieved 5.4 and 4.8 kWh/kg in Cooldry and Mildwet (32% and 30%), and (1,4) achieved 5.9 kWh/kg in Warmdry (39%), closely followed by (1,5) at 6.0 kWh/kg. Hotwet behaved differently. In the asymmetric case, (1,2) gave the lowest SEC (8.7 kWh/kg), followed by (1,7) (8.8 kWh/kg), whereas in the symmetric case (1,7) was lowest (9.7 kWh/kg), followed by (1,2) (10.0 kWh/kg). In both cases, intermediate pairs gave little or no saving in Hotwet. These results suggest a favourable age offset of four to five days in the non-hot scenarios, while in the hot scenario is one day age offset.

Figure 4 shows the HVAC energy of each item of equipment for every age pair. At the reference pair (1,1), total HVAC energy was 1,681 kWh in $Cool_{dry}$, 1,544 kWh in $Mild_{wet}$ and 1,904 kWh in $Warm_{dry}$. The air recirculation unit was the largest single load at (1,1), at 35–43% of the total, but stayed at about 620–670 kWh for every age pair, so it did not explain the differences between them. The differences came mainly from humidification. Moving from (1,1) to (1,5) reduced total HVAC energy consumption by 349 kWh in $Cool_{dry}$, 366 kWh in $Mild_{wet}$ and 558 kWh in $Warm_{dry}$. Likewise, the heater usage had also been reduced leading to the reduced total HVAC energy consumption. The saving at intermediate age pairs therefore came mainly from moisture released by the older larvae, which reduced the need for humidification of the younger larvae.

In $Hot_{wet}$, total HVAC energy at (1,1) was 2,800 kWh, with the humidifier (41%) and the cooler (33%) as the largest loads. Cooling energy was highest at the intermediate pairs, rising from 918 kWh at (1,1) to 1110 kWh at (1,4), which is consistent with the additional metabolic heat adding to the cooling load. At (1,2), cooling stayed close to the reference (888 kWh) while humidification fell by 365 kWh, giving the lowest total of 2,472 kWh. This offers a mechanistic interpretation of the age-pair ranking.

All age pairs met the assumed frass-water criterion (< 50%) in both zone configurations and all four tested scenarios (Table 4c). This criterion therefore did not distinguish between the candidate age pairs.

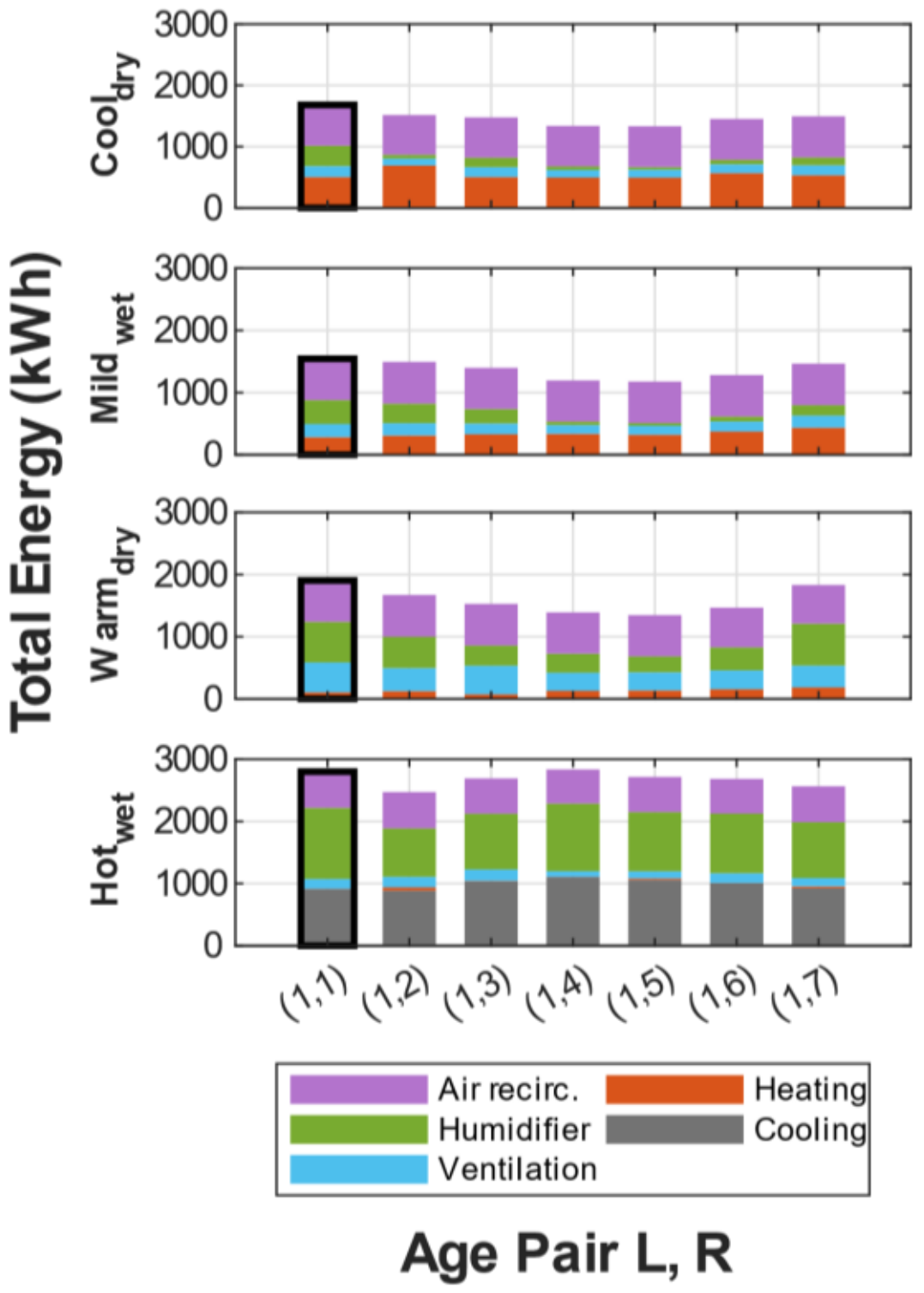

**Figure 4**. HVAC energy breakdown for different age pairs. Total electrical energy consumed by each HVAC component over one 7-day cycle, for each combination of climate scenario and age pair in the asymmetric case (E3). The (1,1) is the reference age pair within each row. The HVAC component shown is heater, ventilator, humidifier and air recirculation unit, with the cooler present only in $Hot_{wet}$.

Age pairing did not replace the need for cooling in $Hot_{wet}$ (Table 5). With pair (1,2), Baseline produced 147 kg of predicted dry biomass and used 1,329 kWh, whereas +Cooler produced 285 kg and used 2,472 kWh. Although SEC differed only modestly (9.0 versus 8.7 kWh/kg), Baseline produced approximately half the biomass and exceeded the assumed frass-water criterion (63.3% versus 36.3%). SEC alone therefore did not capture the operational disadvantage of omitting cooling. The selected asymmetric-configuration pairs were (1,5) for $Cool_{dry}$, $Mild_{wet}$ and $Warm_{dry}$, and (1,2) for $Hot_{wet}$.

**Table 5**. Component trade-off check in the $Hot_{wet}$ scenario: Biomass, HVAC energy, Specific Energy Consumption (SEC), and frass water at harvest for the (1, 2) age pair under +Cooler HVAC and Baseline HVAC.

| Configuration | Biomass (kg) | HVAC energy (kWh) | SEC (kWh/kg) | Frass water (%) |
|---|---|---|---|---|
| **+Cooler** | 284.9 | 2,472.0 | 8.7 | 36.3 |
| **Baseline** | 147.0 | 1,329.0 | 9.0 | 63.3 |

*E4 – Inter-container coupling*

In E4, we compared two coupling techniques in the $Cool_{dry}$ scenario over a 30-day window. Both used the DM controller and the Baseline configuration. The natural coupling technique was two independent containers each running the best asymmetric case age pair from E3, (1,5). The controlled coupling technique was two containers, one holding (1,1) and the other holding (5,5), linked by an actuated duct. The metrics are collected in Table 6.

**Table 6**. E4 performance summary: natural coupling (1,5) versus controlled coupling (1,1)+(5,5) in the $Cool_{dry}$ scenario over a 30-day operating window. Biomass, total energy consumption, Specific Energy Consumption (SEC), frass water and harvest events of one container are reported.

| Metric | Natural Coupling | Controlled Coupling |
|---|---|---|
| **Biomass (kg)** | 2,273 | 2,474 |
| **Total Energy (kWh)** | 14,307 | 14,960 |
| **SEC (kWh/kg)** | 6.29 | 6.05 |
| **Frass water (%)** | 44.3 | 37.1 |
| **Harvest events per container (#)** | 8 | 4 |

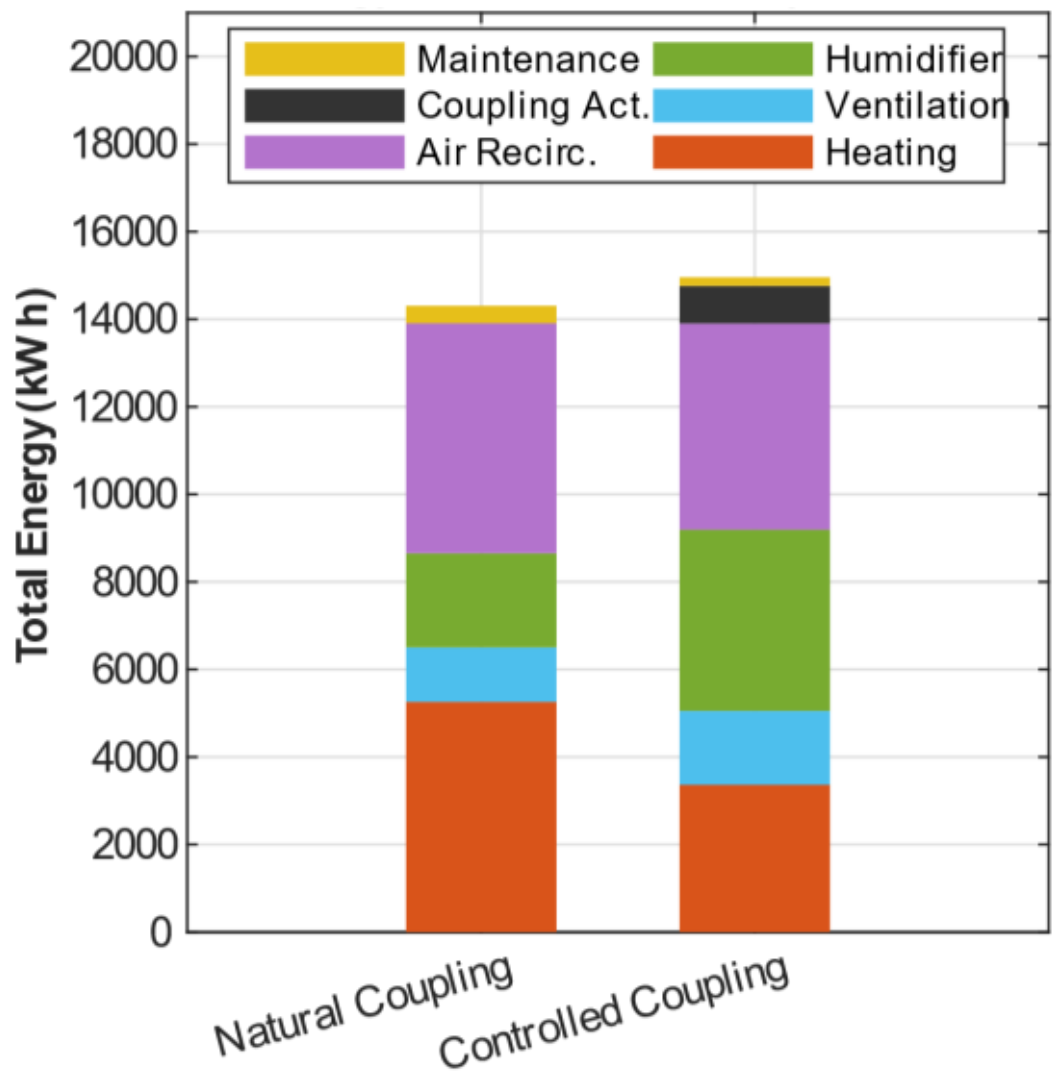

**Figure 5**. Total energy breakdown for coupled containers. Total electrical energy consumed by each HVAC component under natural coupling and controlled coupling in the $Cool_{dry}$ scenario, summed over the 30-day operating window and over both containers (E4). The components shown are heater, ventilator, humidifier and air recirculation unit, with the coupling actuator present only under controlled coupling.

The controlled-coupling arrangement produced approximately 9% more biomass and consumed 4.6% more total HVAC energy than the natural-coupling arrangement (Table 6). The biomass increase exceeded the energy increase, reducing SEC by approximately 3.9%. Both arrangements met the assumed frass-water criterion (44.3% and 37.1%, respectively). The controlled-coupling arrangement therefore reduced the energy consumption per kilogram of harvested dry biomass, but did not reduce total energy consumption.

Figure 5 indicates a redistribution of energy use between actuators. The reported heating demand was approximately 45% lower under controlled coupling, while air-recirculation energy was approximately 11% lower. Air transfer from the older larvae's container could supply heat to the younger larvae's container, with the source and receiving roles changing after harvest. Whole-container harvesting also reduced the frequency of control interruptions and recovery periods.

Humidifier and ventilator usage increased to condition the transferred air. The air arriving from the neighbouring container carried useful heat but was not at the target humidity, so it had to be re-conditioned before entering the receiving container. Additionally, a small dedicated coupling-actuator load was added, which was why total HVAC energy was 4.6% higher despite the large heating reduction.

In summary, the controlled-coupling arrangement achieved higher predicted biomass, lower SEC and fewer reported harvest events. These benefits were accompanied by higher total energy consumption and additional coupling hardware. The small SEC improvement does not, by itself, establish a financial or environmental advantage for the coupled arrangement.

*Limitations and future work*

This study is entirely simulation-based. Laboratory validation of the underlying biological model does not validate the container-scale heat and mass transfer model or the predicted industrial performance. The four experiments explored a broad design space, but used different subsets of the eight climate scenarios. Their results identify candidate operating strategies for further testing; quantitative savings and hardware selections require validation against independent container-farm measurements.

The biological parameters were derived from a single calibrated source, and the imposed mismatch involved only two parameters. This is a limited uncertainty test, not a comprehensive robustness analysis. Feed composition, larval density, microbial heat generation, transfer coefficients, actuator efficiency and objective weights could change the rankings and predicted benefits. Real-time biomass and substrate-moisture information was assumed available, although these quantities are difficult to measure directly. State estimation is addressed in separate work (Mansour et al., 2026; Tafat et al., 2026). Actuator wear and financial maintenance costs were not modelled. Biosecurity risks associated with inter-container air exchange were also not quantified and require assessment before deployment. In E4, the effects of coupling, age allocation and harvest organisation were not separated; comparisons using matched initial and final inventories and identical harvesting schedules would help isolate the coupling benefit.

A further limitation concerns the control framework. The two coupled containers were controlled by a single centralised hierarchical controller. This was tractable for two containers. Because the size of the combined optimisation problem grows with each added container, computation time is expected to increase with the number of containers, and a centralised approach may become impractical for a large facility. A distributed implementation, in which each container is controlled locally and exchanges limited information with its neighbours, is a possible alternative. Similarly, E4 showed a benefit of coupling for two containers, but how this benefit depends on network size, topology and age composition was not studied. Distributed control and larger coupled networks are left for future work.

## 4. Conclusions

The simulations investigated whether predictive climate control and reuse of internally generated heat and moisture could improve the energy intensity of container-based BSFL bioconversion. The results support four provisional operating and design insights within the tested model and scenarios:

- Dynamic climate setpoints achieved lower SEC than a single fixed setpoint pair in all four scenarios.
- HVAC selection depended on the climate scenario and control strategy; active cooling was needed in the two hot scenarios, whereas active dehumidification was not retained under the tested settings.
- Staggered larval age pairs could reduce heating and humidification demand in the tested non-hot scenarios, while the preferred age offset differed in hot and wet climate conditions.
- In the cold and dry climate scenario, controlled inter-container coupling combined with age separation and changed harvest organisation increased biomass and modestly reduced SEC, despite increasing absolute energy consumption.

The framework provides a tool for exploring candidate climate-control strategies and container arrangements before targeted experimental testing. Its current role is hypothesis generation and comparative simulation, rather than validated industrial component sizing. Independent container-scale measurements, sensitivity analysis and comparisons that isolate coupling from age allocation and harvest timing are needed before quantitative recommendations can guide investment or operation. Extensions to other insect species and larger container networks would require species-specific calibration and further validation.

## Supplementary material

The Supplementary material contains the model equations and the larval development rate as a function of medium temperature (Section S1, Figure S1), the control formulations (Section S2), and the model parameters (Table S1), the Digital Twin parameter changes (Table S2), the controller settings and the forecast error definition (Section S3).

Supplementary Material

# Metabolic heat and moisture reuse with climate-adaptive control for energy-efficient black soldier fly bioconversion in modular container farms

S. Saravanan[1,a], P. Sauerteig[1,b], M. Gold[2,3,c], S. Streif[1,4,d,*]

[1] Professorship Automatic Control and System Dynamics, Technische Universität Chemnitz, 09126 Chemnitz, Germany (e-mail: **sriram.saravanan@etit.tu-chemnitz.de**; **philipp.sauerteig@etit.tu-chemnitz.de**; **stefan.streif@etit.tu-chemnitz.de**).
[2] ETH Zurich, Sustainable Food Processing, Zurich, Switzerland
[3] Reploid Group AG, Wels, Austria (e-mail: **gold@reploid.eu**)
[4] Fraunhofer Institute for Molecular Biology and Applied Ecology IME, Department of Bioresources, Giessen, Germany.

*Corresponding author: **stefan.streif@etit.tu-chemnitz.de**.

[a] ORCID: https://orcid.org/0009-0007-6113-0983
[b] ORCID: https://orcid.org/0000-0002-5975-5147
[c] ORCID: https://orcid.org/0000-0001-7285-6164
[d] ORCID: https://orcid.org/0000-0002-3398-1226

## S1. Detailed Container Model

This section provides the complete set of state variables, control inputs, ODEs, and flux expressions for the detailed container model summarised in Materials and Methods, together with the simplified-model derivation and the coupling equations.

*Notation*

Throughout this supplementary material, we use T for temperature, H for humidity, C for $CO_2$ concentration, and O for $O_2$ concentration, with subscripts indicating the subsystem. Exchange fluxes are denoted by $\phi$, with double subscripts of the form 'quantity, subsystem pair' (e.g., $\phi_{Q,con-try}$ is a heat flux from container to tray). The order of the two subsystems in a subscript is a naming convention only and does not indicate the direction of flow. The sign of each flux follows from its defining expression and from the equation in which it appears. Physical and actuator constants are denoted by k with descriptive subscripts.

Each subsystem is described by three kinds of variables. The states x are quantities that change over time as the system operates (for example, container air temperature, medium water content, or larval biomass). The inputs u are the quantities that can be set directly to influence how the states evolve; the actuator commands provided by the HVAC equipment and the air-recirculation hardware are the inputs. The disturbances are external quantities that also influence the states but cannot be controlled. In this model, the disturbances are the outdoor weather conditions:

$$d = [T_{out}, H_{out}, C_{out}, O_{out}]^T.$$

The outdoor disturbances act on the container through three pathways: passive heat conduction through the walls, air exchange when the door is open, and air exchange when the ventilator draws outside air into the pipe. For each state, an ODE describes how fast that state changes given the current values of the states, inputs, and disturbances.

*State and input vectors*

*Larval growth subsystem*

The larval submodel has five states per zone. It is collected in

$$x_{lrv} = \left[T_{med}, B_{dry}, W_{med}, N_{med}, N_{exc}\right]^T$$

where$T_{med}$ is the medium temperature, $B_{dry}$ is the average dry biomass per larva, $W_{med}$ is the medium water content, $N_{med}$ is the unconsumed nutrient mass, and $N_{exc}$ is the accumulated excreta.

*Tray microclimate subsystem*

The tray submodel has four states per zone. It is collected in

$$x_{try} = \left[T_{try}, H_{try}, C_{try}, O_{try}\right]^T.$$

*Climate container subsystem*

Each zone $z \in \{l, r\}$ has four states. They are collected in

$$x_{con,z} = [T_{con,z}, H_{con,z}, C_{con,z}, O_{con,z}]^T$$

*Recirculation pipe subsystem*

The pipe submodel has four states. They are collected in

$$x_{pipe} = [T_{pipe}, H_{pipe}, C_{pipe}, O_{pipe}]^T$$

*Detailed container state and input*

Collecting everything, the detailed container state and input vectors are

$$x_{fc} = [x_{con,l}^T, x_{try,l}^T, x_{lrv,l}^T, x_{con,r}^T, x_{try,r}^T, x_{lrv,r}^T, x_{pipe}^T]^T$$

$$u_{fc} = [u_{pipe}^T, u_{con}^T,]^T$$

where $u_{con} = [u_A, u_{door}]^T$ collects the recirculation-flow input $u_A \in [0, 1]$ and the binary door input $u_{door} \in \{0, 1\}$, and $u_{pipe} = [u_T, u_V, u_H, u_D]^T$ collects the heater/cooler, ventilator, humidifier, and dehumidifier inputs.

The system dynamics are written as

$$\dot{x}_{fc} = f_{full}(x_{fc}, u_{fc}, d).$$

*Larval subsystem ODEs*

The medium temperature evolves as

$$k_{c,med}\,\dot{T}_{med} = \phi_{Q,med} - \phi_{Q,try-med} - \phi_{Q,lat-med}$$

where $k_{c,med}$ is the heat capacity of the medium, $\phi_{Q,med}$ collects the metabolic heat released by the larvae and the microbiome, and $\phi_{Q,lat-med}$ is the latent heat flux. The dry biomass per larva follows the energy balance between assimilation and maintenance:

$$\dot{B}_{dry} = \phi_{assim} - \phi_{maint}$$

with $\phi_{assim}$ and $\phi_{maint}$ given by the rate laws described in Padmanabha et al. (2020). Both fluxes depend on the medium temperature through the larval development rate, which follows the Logan-10 model. Figure S.1 shows that the rate rises with medium temperature to a peak at the developmental optimum of about 35.1 °C, then falls sharply toward zero at $T_{max} =$ 39.5 °C.

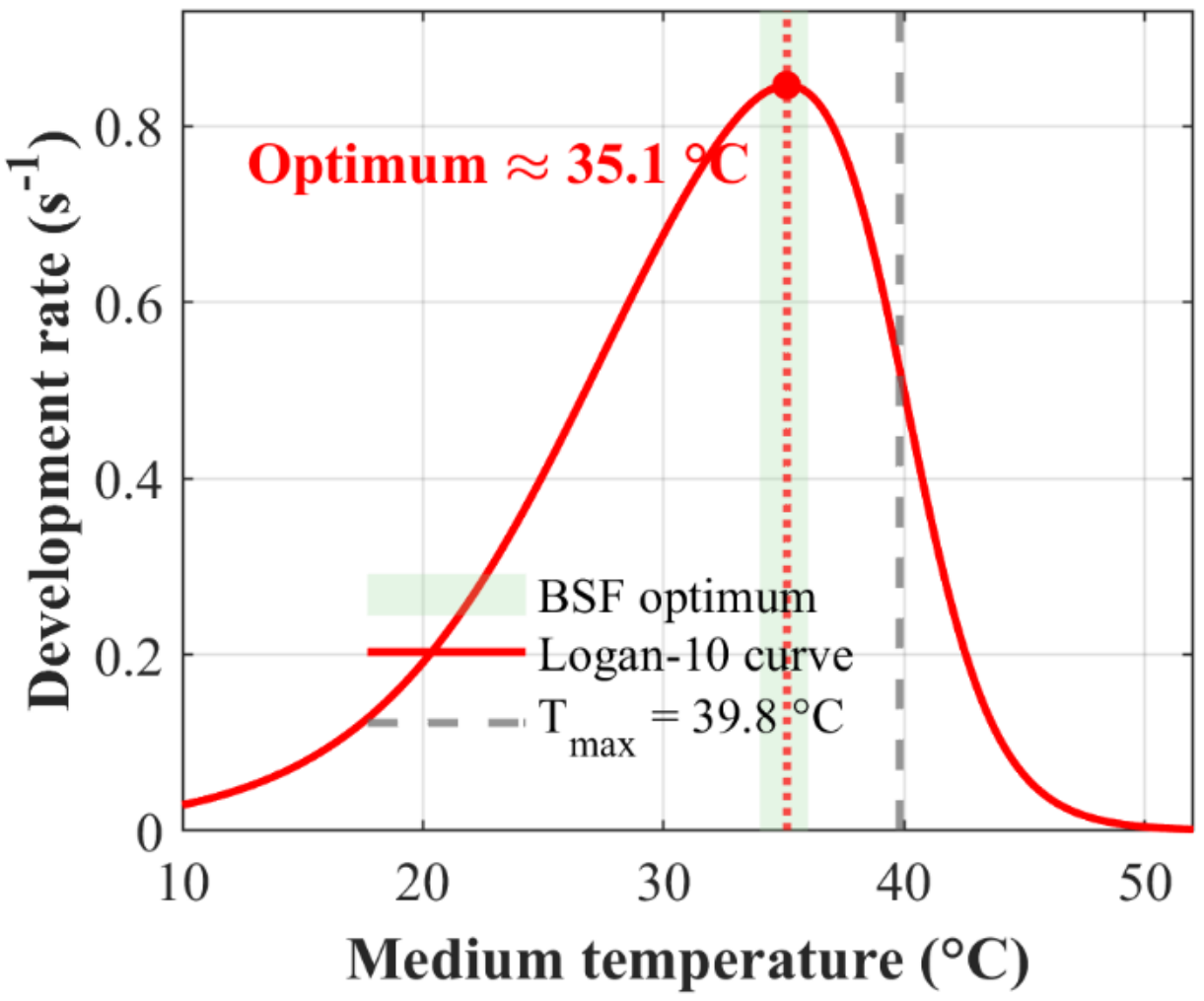

Supplementary Figure S1. Larval development rate as a function of medium temperature under the Logan-10 model, with the developmental optimum near 35.1 °C and $T_{max} = 39.5\,°C$.

The medium water content is depleted by larval consumption $\phi_{W,subj}$ and by evaporation $\phi_{H,med}$:

$$\dot{W}_{med} = -\phi_{W,subj} - \phi_{H,med}$$

The nutrient state decreases through ingestion and is partially returned as excreta:

$$\dot{N}_{med} = k_{lrv,cnt}(-\phi_{ing} + \phi_{exc})$$

The excreta state accumulates the residual fraction:

$$\dot{N}_{exc} = k_{lrv,cnt} k_{exc,med}\, \phi_{exc}$$

where $k_{lrv,cnt}$ is the number of larvae in the tray and $k_{exc,med}$ is the excreted fraction of ingested feed.

*Tray microclimate ODEs*

For each generic state $\xi \in \{T, H, C, O\}$, the container-tray exchange combines forced and natural convection:

$$\phi_{\xi,con-try} = \kappa_\xi(\,|u_A| k_{a-t,ex,f} + k_{a-t,ex,n})(\xi_{try} - \xi_{con})$$

with $\kappa_T = k_{T,air}$ (volumetric heat capacity of air) and $\kappa_\xi$=1 for $H, C$ and $O$. The forced and natural exchange coefficients are $k_{a-t,ex,f}$ and $k_{a-t,ex,n}$.

The tray state ODEs are

$$\begin{aligned} k_{c,try}\dot{T}_{try} &= (\phi_{Q,try-med} - \phi_{Q,con-try}), \\ k_{v,try}\dot{H}_{try} &= (\phi_{H,med} - \phi_{H,con-try}), \\ k_{v,try}\dot{C}_{try} &= (\phi_{C,sub} - \phi_{C,con-try}), \\ k_{v,try}\dot{O}_{try} &= (\phi_{O,sub} - \phi_{O,con-try}), \end{aligned}$$

where $k_{c,try}$ is the tray heat capacity, $k_{v,try}$ is the tray air volume, and $\phi_{C,sub}$, $\phi_{O,sub}$ are the lumped $CO_2$ and $O_2$ fluxes from larval and microbial respiration.

*Climate container ODEs*

The temperature dynamics of zone z combine wall conduction, door exchange, zone-to-zone exchange, and the container-tray exchange:

$$k_{c,z}\dot{T}_{con,z} = \phi_{Q,wall,z} + \phi_{Q,door,z} + \phi_{Q,z-z,z} + \phi_{Q,con-try}$$

where $k_{c,z}$ is the zone heat capacity. The four fluxes, from left to right, are: passive heat conduction through the walls,

$$\phi_{Q,wall,z} = k_{T,cond}\left(T_{out} - T_{con,z}\right)$$

heat exchange through the door when open,

$$\phi_{Q,door,z} = u_{door}k_{T,air}k_{\dot{u},door}\left(T_{out} - T_{con,z}\right)$$

zone-to-zone convective exchange via the recirculation loop; and the container-tray exchange already given above. The zone-to-zone flux uses the positive and negative parts of the recirculation input,

$$u_A^+ = \max(u_A, 0)\,,$$
$$u_A^- = \min(u_A, 0),$$

to select the active flow direction. For the left zone,

$$\phi_{Q,z-z,l} = k_{T,air}k_{a-t,ex,f}\left[u_A^+\left(T_{pipe} - T_{con,l}\right) + u_A^-\left(T_{con,r} - T_{con,l}\right)\right] + k_{a-t,ex,n}\left(T_{con,r} - T_{con,l}\right),$$

where the first term reflects fresh air arriving from the pipe when flow is left-to-right, the second term is the air exchange with the right zone when the direction is reversed, and the third is the natural convection component. The right-zone expression is symmetric:

$$\phi_{Q,z-z,r} = k_{T,air}k_{a-t,ex,f}\left[u_A^+\left(T_{con,l} - T_{con,r}\right) + u_A^-\left(T_{pipe} - T_{con,r}\right)\right] + k_{a-t,ex,n}\left(T_{con,l} - T_{con,r}\right),$$

The humidity, $CO_2$, and $O_2$ dynamics of zone z follow the same template with the wall conduction term dropped (the container is assumed vapour- and gas-tight along the walls):

$$k_{v,z}\dot{H}_{con,z} = \phi_{H,con-try} + \phi_{H,z-z,z} + \phi_{H,door,z},$$
$$k_{v,z}\dot{C}_{con,z} = \phi_{C,con-try} + \phi_{C,z-z,z} + \phi_{C,door,z},$$
$$k_{v,z}\dot{O}_{con,z} = \phi_{O,con-try} + \phi_{O,z-z,z} + \phi_{O,door,z},$$

where $k_{v,z}$ is the zone air volume. The door and zone-to-zone fluxes for humidity, $CO_2$, and $O_2$ have the same structural form as the temperature case, with T replaced by the corresponding state and the $k_{T,air}$ factor dropped (it appears only for heat).

*Recirculation pipe ODEs*

The pipe temperature dynamics combine the heater/cooler flux, the ventilator flux, and the container-pipe flux:

$$k_{c,pipe}\dot{T}_{pipe} = \phi_{heat} + \phi_{Q,vent} + \phi_{Q,p-con}$$

with the heater/cooler flux

$$\phi_{heat} = u_T k_{heat,max}$$

where $k_{heat,max}$ is the maximum thermal power in either direction.

The ventilator flux

$$\phi_{Q,vent} = u_V k_{T,air} k_{\dot{V},max}(T_{out} - T_{pipe})$$

where $k_{\dot{V},max}$ is the maximum air flow rate and the container-pipe heat exchange flux

$$\phi_{Q,p-con} = k_{T,air} k_{a-t,ex,f} \left[u_A^+(T_{con,r} - T_{pipe}) + u_A^-(T_{con,l} - T_{pipe})\right],$$

which reflects that air enters the pipe from the right zone when the airflow is left-to-right and from the left zone when reversed.

The pipe humidity dynamics combine the humidifier flux, the dehumidifier flux, the ventilator flux, and the container-pipe exchange:

$$k_{v,pipe}\dot{H}_{pipe} = \phi_{H,hum} + \phi_{H,dehum} + \phi_{H,vent} + \phi_{H,p-con}$$

The humidifier injects water vapour at a rate proportional to the gap between the current pipe humidity and its temperature-dependent saturation limit $H_{sat}(T_{pipe})$:

$$\phi_{H,hum} = u_H k_{\dot{u},H}( H_{sat}(T_{pipe}) - H_{pipe})$$

where $k_{\dot{u},H}$ is the maximum humidification rate. The dehumidifier removes water vapour at a rate proportional to the current pipe humidity:

$$\phi_{H,dehum} = -u_D k_{\dot{u},D} H_{pipe},$$

where $k_{\dot{u},D}$ is the maximum dehumidification rate. The ventilator and container-pipe humidity fluxes have the same structural form as the temperature case with $k_{T,air}$ dropped.

The $CO_2$ and $O_2$ dynamics of the pipe combine the ventilator and container-pipe fluxes only:

$$k_{v,pipe}\dot{C}_{pipe} = \phi_{C,vent} + \phi_{C,p-con},$$
$$k_{v,pipe}\dot{O}_{pipe} = \phi_{O,vent} + \phi_{O,p-con}.$$

*Simplified model equations*

The simplified model exploits the timescale difference between the climate and the biology. On the slow timescale of the larvae, the container climate has effectively reached its steady state for any given control input. We replace the climate differential equations with algebraic offsets relative to the outside environment:

$$T_{con,z} = T_{out} + u_{\Delta T},$$

$$H_{con,z} = H_{out} + u_{\Delta H},$$

where $u_{\Delta T}$ and $u_{\Delta H}$ are temperature and humidity offsets that the HVAC is assumed to provide. The simplified model assumes a single climate setpoint for the whole container, applied to both zones. This is consistent with the physical arrangement, in which the two zones share the same container air, so their climates cannot be set independently. $CO_2$ and $O_2$ are taken to track the outside in the simplified model, since fresh-air ventilation in normal operation keeps them within feasible bands. The tray, pipe, and HVAC submodels are subsumed into this static map. The simplified control vector is

$$u_{red} = [u_{\Delta T}, u_{\Delta H}]^T,$$

and the simplified dynamics are written compactly as

$$\dot{x}_{red} = f_{red}(x_{red}, u_{red}, d).$$

The simplified model trades realism (no container thermal inertia, no per-actuator energy accounting) for computational tractability at the planning level.

*Coupling equations*

*Coupling in the simplified model*

Because the simplified model has no climate states, coupling cannot be added at the state level. Instead, it acts directly on the algebraic climate offsets. For containers i and j,

$$T_{con,z,i} = T_{out} + u_{\Delta T,i} + \gamma_{ij}(T_{con,z,j} - T_{con,z,i}),$$

$$H_{con,z,i} = H_{out} + u_{\Delta H,i} + \gamma_{ij}(H_{con,z,j} - H_{con,z,i}),$$

where the non-negative coupling strength $\gamma$ encodes how strongly the two containers exchange heat and humidity; $\gamma = 0$ recovers the decoupled case, and as $\gamma$ increases, the two containers' temperatures equilibrate.

In principle, $\gamma$ can take any non-negative value, but physically it is limited by the cross-sectional area of the coupling duct and the maximum throughput of its fan. We therefore bound it as $\gamma \in [0, \gamma_{max}]$, where $\gamma_{max}$ is the maximum achievable coupling strength when the coupling actuator is fully open. $CO_2$ and $O_2$ remain uncoupled in the simplified model. The input vector of container i is extended to include the coupling strength:

$$u_{red,i}^{coup} = \left[u_{red,i}^T, \gamma\right]^T.$$

Solving the simplified-model coupling equations explicitly for two containers yields the closed-form solution

$$T_{con,z,1} = T_{out} + \frac{(1 + \gamma)u_{\Delta T}, 1 + \gamma\, u_{\Delta T}, 2}{(1 + 2\gamma)},$$

$$T_{con}, z, 2 = T_{out} + \frac{\gamma\, u_{\Delta T}, 1 + (1 + \gamma)u_{\Delta T}, 2}{(1 + 2\gamma)}.$$

The humidity expressions are analogous.

*Coupling in the detailed model*

In the detailed model, the two containers exchange heat, humidity, $CO_2$, and $O_2$ directly through their pipes via a controllable coupling actuator $u_C \in [0, 1]$. The pipe temperature ODE of container i gains an additional coupling flux

$$\phi_{Q,coup,i} = u_C k_{T,air}\, k_{\dot{V},coup}\left(T_{pipe,j} - T_{pipe,i}\right)$$

where $j \neq i$ is the other container and $k_{\dot{V},coup}$ is the maximum coupling air flow rate. Analogous fluxes $\phi_{H,coup,i}$, $\phi_{C,coup,i}$, $\phi_{O,coup,i}$ are added to the humidity, $CO_2$, and $O_2$ pipe equations. To stay consistent with the simplified-model abstraction, $u_C = \frac{\gamma}{\gamma_{max}}$, so the coupling strength chosen by the upper-level planner directly maps to the actuator opening commanded at the lower level. The input vector of container i becomes

$$u_{fc,i}^{coup} = \left[u_{fc,i}^{T}, u_C\right]^{T}.$$

*Time discretisation*

For both the detailed and simplified models, the continuous-time dynamics are discretised using the implicit Euler scheme with sampling time $\Delta t$ (Biswas et al., 2013). We denote the discretised dynamics of the detailed, simplified, detailed-with-coupling, and simplified-with-coupling cases by $F_{full}$, $F_{red}$, $F_{full}^{coup}$, and $F_{red}^{coup}$ respectively.

## S2. Detailed control formulations

*Upper-level OCP*

The upper-level objective combines three terms: maximising the dry biomass at harvest, minimising the frass water at the end of the bioconversion cycle, and penalising actuator effort. Combining and weighting these gives

$$J_{up} = -\alpha_1 B_{dry}(N_{up}) * k_{lrv,cnt} + \alpha_2 W_{med}(N_{up}) + \Sigma_{k=0}^{N_{up}-1} \|u(k)\|_2^R, \tag{1}$$

where $\alpha_1, \alpha_2 \geq 0$ are weights, $B_{dry}(N_{up})$ is the estimated dry biomass of both zones at the harvest and R scales the individual control inputs. The simplified model represents the HVAC system as temperature and humidity offsets and has no per-actuator energy accounting, so electrical energy cannot be computed at the upper level. Penalising the offsets therefore acts as a surrogate for energy use. The residual substrate water mass enters the objective as a penalty rather than as a terminal constraint, because a hard constraint on substrate water over a 7-day horizon can make the problem infeasible. The same weights are used in every configuration and scenario, so these choices affect all results in the same way and do not change the comparisons between them. The upper-level optimal control problem, solved once per day over a prediction horizon of $N_{up}$ steps, is written as follows:

$$\min_{u} \; J_{up} \tag{2a}$$

$$\text{s.t.} \;\; x(0) = x_0, \tag{2b}$$

$$x(k+1) = F_{red}\big(x(k), u(k), d(k)\big), \tag{2c}$$

$$\big\|u(k) - u(k-1)\big\|_{\infty} \leq \Delta_u^{up}, \tag{2d}$$

$$x(k) \in X, \tag{2e}$$

$$u(k) \in U, \tag{2f}$$

where $J_{up}$ is the upper-level objective defined in (1) and $F_{red}$ is the discretised simplified model. Constraint (2b) fixes the initial predicted state to the current measured state. Constraint (2c) propagates the predicted state forward. Constraint (2d) limits how aggressively the climate setpoint may change between consecutive upper-level steps. The sets X and U encode larval thermal-tolerance and HVAC operation bounds. The disturbance $d(k)$ is the predicted outdoor climate over the planning horizon, assumed known from weather forecasts.

*Harvest-point optimisation*

Harvesting at candidate time $t_h^m$ means terminating the current bioconversion cycle at that step and starting a new one immediately afterwards. Fresh larvae for bioconversion are prepared outside the container in advance, so the switchover only requires moving trays in and out. We allow a one-hour idle window during which no control inputs are applied, since the container door is open. The initial state of the new bioconversion cycle may differ from the previous one and is assumed to be known. In principle the harvest time is itself a decision variable. We therefore evaluate the OCP over a finite set of candidate harvest times $T_h = \{t_h^1, \dots, t_h^M\}$ and select the candidate with the best objective value.

*Lower-level MPC*

The lower level tracks the upper-level climate reference $y^{ref} = \left(T_{con,l}^{ref}, H_{con,l}^{ref}, T_{con,r}^{ref}, H_{con,r}^{ref}\right)^T$, denoting the controlled container outputs by $y = \left(T_{con,l}, H_{con,l}, T_{con,r}, H_{con,r}\right)^T$. The objective combines tracking error and actuator effort:

$$J_{lo} = \Sigma_{k=0}^{N_{lo}-1} \left\|y(k) - y^{ref}(k)\right\|_Q^2 + \|u_{fc}(k)\|_{R_{lo}}^2, \qquad (3)$$

where Q weighs deviations from the reference and $R_{lo}$ weighs the control effort. The lower-level MPC, solved every $\Delta t_{lo}$ over a prediction horizon of $N_{lo}$ steps, is

$$\begin{aligned}
&\min_u J_{lo},\\
&\text{s.t. } x(0) = x_0,\\
&\quad x(k+1) = F_{full}\big(x(k), u(k), d(k)\big),\\
&\quad \big\|u(k) - u(k-1)\big\|_\infty \le \Delta_u^{lo},\\
&\quad x(k) \in X,\\
&\quad u(k) \in U,
\end{aligned}$$

where $J_{lo}$ is the lower-level objective function defined in (3), $u = u_{fc}$ is the vector of HVAC control inputs, and $F_{full}$ is the discretised detailed model. The structure mirrors the upper-level OCP, with the state set X additionally enforcing a $CO_2$ concentration bound that lies outside the scope of the simplified model. The disturbance $d(k)$ is the predicted outdoor climate over the next $N_{lo}$ steps.

*Coupled-container formulation*

For two coupled containers, the pair is treated as a single system at both levels. At the upper level, the OCP minimises the sum of both containers' objectives over the coupled simplified-model dynamics, and the coupling strength $\gamma \in [0, \gamma_{max}]$ becomes an additional decision variable. The harvest times of the two containers are optimised independently within the same problem.

At the lower level, a single MPC tracks the joint setpoint reference $\tilde{y}^{ref} = [y_1^{ref}, y_2^{ref}]$ produced by the upper level, together with the coupling-actuator reference $u_C^{ref}$. The lower-level objective gains an additional term penalising deviation of the coupling actuator from its reference, weighted by $R_C$:

$$J_{coup} = \Sigma_{k=0}^{N_{lo}-1} \left\|y(k) - \tilde{y}^{ref}(k)\right\|_Q^2 + \|u_{fc}(k)\|_{R_{lo}}^2 + \left\|u_C(k) - u_C^{ref}(k)\right\|_{R_C}^2.$$

Rate and saturation constraints apply individually to each container, and the coupling actuator carries its own bounds.

## S3. Numerical values and configuration

This section collects the numerical values used in the prediction and simulation models, the controller configuration, the forecast noise levels, and the performance-metric expressions.

*Model parameters*

The detailed and simplified models share a single parameter set. Parameters that were re-scaled for the container-farm setting, together with those introduced in this work, are listed in Table 1. All parameters not listed there took the values reported in Padmanabha et al., 2020, which are not repeated here.

Supplementary Table S1. Prediction model parameters

| Parameter | Value | Unit |
|---|---|---|
| $k_{A,con}$ | 94.32 | $m^2$ |
| $k_{V,con}$ | 32.525 | $m^3$ |
| $k_{A,try}$ | 72.8 | $m^2$ |
| $k_{V,try}$ | 21.6 | $m^3$ |
| $k_{A,pipe}$ | 4.32 | $m^2$ |
| $k_{V,pipe}$ | 0.4 | $m^3$ |
| $k_{A,med}$ | 108 | $m^2$ |
| $k_{V,med}$ | 10.8 | $m^3$ |
| $k_{T,air}$ | 1,211.3 | $J/m^3K$ |
| $k_{a-t,ex,f}$ | 1.4 | $m^3/s$ |
| $k_{a-t,ex,n}$ | 0.1 | $m^3/s$ |
| $k_{heat,max}$ | 7,000 | W |
| $k_{V,max}$ | 1.4 | $m^3/s$ |
| $k_{\dot{u},H}$ | 15.5 | $Kg/m^3s$ |
| $\eta_{hum}$ | 0.9 | - |
| $\eta_{dehum}$ | 0.9 | - |
| $\gamma_{max}$ | 3.5 | - |
| $k_{\dot{u},D}$ | 15.5 | $Kg/m^3s$ |
| $k_{lrv,cnt}$ | 14,000 | - |
| $P_T$ | 7 | kW |
| $P_V$ | 4 | kW |
| $P_A$ | 4 | kW |
| $P_C$ | 4 | kW |

*Energy calculation*

The humidifier and dehumidifier energies in Equation 1 of the main text were calculated from the water vapour added to or removed from the recirculation pipe (Jo et al., 2017):

$$E_{hum} = \sum_{k=1}^{N_{sim}} \frac{u_H(k) k_{u,\dot{H}} l_v \max\left(H_{sat}\left(T_{pipe}(k)\right) - H_{pipe}(k), 0\right) \Delta t}{\eta_{hum}}$$

$$E_{dehum} = \sum_{k=1}^{N_{sim}} \frac{u_D(k) k_{u,\dot{D}} l_v H_{pipe}(k) \Delta t}{\eta_{dehum}}$$

where $l_v$ is the latent heat of vaporisation, $H_{sat}$ is the saturation humidity at the pipe temperature, and $\eta_{hum}$ and $\eta_{dehum}$ are the efficiencies of the humidifier and dehumidifier.

*Digital Twin parameter variation*

The Digital Twin uses the same parameter set as the prediction model, except for two larval growth parameters whose values are perturbed relative to their nominal values:

Supplementary Table S2. Biological parameters of the larvae growth model that are varied with respect to the source.

| Parameter | Values (Padmanabha et al., 2020) | Altered Values |
|---|---|---|
| maximum moisture threshold: $k_{lrv,moist,max}$ | 1.16 (g/s) | 1.1(g/s) |
| maximum growth rate threshold: $k_{lrv,rate,max}$ | 1.25 (1/s) | 1.15 (1/s) |

Both are reduced relative to nominal. This perturbation mimics the biological uncertainty a controller would face in a real industrial setting, where larval growth parameters are known to vary substantially with environmental conditions, feed composition, and substrate moisture (Padmanabha et al., 2020).

*Controller configuration*

*Upper-level OCP*

The objective weights are $\alpha_1 = 2$, $\alpha_2 = 1$, and $R = \text{diag}([10^{-5}, 10^{-3}])$. The prediction horizon for the upper-level OCP is $N_{up}$ = 168 and the sampling time is $\Delta t_{up}$ = 1 hour. State constraints per zone are

- $T_{con,z} \in [5, 40]$ °C;
- $H_{con,z} \in [10, 95]$ %;
- $C_{con,z} \in [0, 5000]$ ppm.

Input constraints are $u_{\Delta T} \in [0, 30]$°C and $u_{\Delta H} \in [0, 0.02]$ kg/m$^3$ for HVAC configurations without cooling and without dehumidification. Input constraints for HVAC configuration with cooling and dehumidification are $u_{\Delta T} \in [-30, 30]$°C and $u_{\Delta H} \in [-0.02, 0.02]$ kg/m$^3$. The rate-of-change bound is $\Delta_u^{up} = 3$ K for the temperature offset and 10% for the humidity offset between consecutive upper-level steps.

*Lower-level MPC*

The objective weight matrices are $Q = 10\,\text{diag}([1, 1,1,1])$ for climate tracking error and $R_{lo} = 10^{-3}\,\text{diag}([10, 1, 10, 1, 10,1])$ for the container and pipe inputs. The prediction horizon for the lower-level MPC is $N_{lo}$ = 4 and the sampling time is $\Delta t_{lo}$ = 15 minutes. The $CO_2$ state constraint $C_{con,z} \in [0, 5000]$ ppm from the upper level is retained. The heater/cooler is bounded as $u_T \in [-1,1]$ with negative values indicating cooling. All other inputs are bounded between $u_{fc} \in [0, 1]$. The rate-of-change bound is $\Delta_u^{lo} = 10\%$, applied component-wise to all inputs in normalised units.

*Coupled-container extension*

For the two-container coupling case, the weights are repeated accordingly for the second container. The coupling strength $\gamma$ becomes an additional decision variable in the upper-level OCP, bounded as $\gamma \in [0, 3.5]$ and penalised in the objective with weight $R_C^{up} = 10^{-7}$. The corresponding coupling-actuator tracking term in the lower-level MPC carries weight $R_C^{lo} = 10$.

The cost weights at both levels are manually tuned to obtain qualitatively reasonable trade-offs among biomass yield, control effort, and tracking quality. Constraint values that deviate from those listed here in a specific experiment are noted in the corresponding experimental design subsection of the main text.

*HVAC configuration variants*

The four HVAC configurations compared in E2 differ only in which actuators are available. An actuator absent from a configuration is disabled by fixing its input to zero for the whole simulation. It therefore contributes neither actuation to the model nor electrical energy. All remaining bounds, weights and rate limits are unchanged. In the baseline configuration the cooler and the dehumidifier are disabled, in +Cooler only the dehumidifier, in +Dehum only the cooler, and in +Cooler+Dehum all actuators are available.

*Weather forecast*

The two control levels do not have access to the actual weather d_sim. Each instead receives a forecast, generated by adding a random error to d_sim:

$$d_{sim} = \text{Inness et al.}, 2019,$$

$$d_{lo} = d_{sim} + \eta_{lo},$$

$$d_{up} = d_{sim} + \eta_{up},$$

where $d_{up}$ is the upper-level forecast, $d_{lo}$ is the lower-level forecast, and $\eta_{lo}$ is a normal distribution with zero mean and variance $\sigma_{lo} = 0.05 * (\max(d_{sim}) - \min(d_{sim}))$ , $\eta_{up}$ is a normal distribution with zero mean and variance $\sigma_{up} = 0.10 * (\max(d_{sim}) - \min(d_{sim}))$ are the noise processes acting on the respective level. The upper-level OCP sees the coarse signal $d_{up}$, consistent with a multi-day weather forecast that does not resolve short-term fluctuations. The lower-level MPC sees the refined forecast $d_{lo}$. The Digital Twin evolves under the actual weather d_sim, so the controller always acts on information that differs from the conditions the container experiences.

The forecast error is included so that neither control level has perfect knowledge of the weather, as in practice. It is not used to assess robustness to forecast uncertainty. All results use a single realisation of the error, generated with a fixed seed, so that all controllers and configurations within an experiment receive the same forecasts.